\documentclass[twocolumn,showpacs,preprintnumbers,prd,nofootinbib]{revtex4-1}
\usepackage{CJKutf8}
\usepackage{feynman}
\usepackage{graphicx,amsmath,amssymb,xcolor,hyperref}
\usepackage{slashed}
\usepackage{cancel}
\newcommand{\tabincell}[2]{\begin{tabular}{@{}#1@{}}#2\end{tabular}}

\begin{document}

\author{Yi Chung}
\email{yichungphys@gmail.com}
\affiliation{Max-Planck-Institut f\"ur Kernphysik, Saupfercheckweg 1, 69117 Heidelberg, Germany}

\title{Generating the Dark Matter mass from the QCD vacuum:\\A new approach to the mass aspect of the Dark Matter-Baryon coincidence problem}

\begin{abstract}
The comparable abundances of dark matter and baryons imply a deep connection between the dark sector and the QCD sector. In asymmetric dark matter models, the number densities of both sectors are ensured to be similar. However, a complete solution should also include a mechanism relating their masses. In this study, we explore a new mechanism where an $\mathcal{O}(\Lambda_{\rm QCD})$ dark matter mass arises from the QCD vacuum. The idea is realized by a strongly coupled chiral dark sector, featuring a light composite fermion as the dark matter. Its mass is generated through the vacuum misalignment induced by QCD, ensuring the mass to be comparable with baryons. The model also features additional light bosons, including GeV-scale dark pions and a composite QCD axion.
\end{abstract}

\maketitle

\section{Introduction}

The Standard Model (SM) of particle physics successfully describes all known elementary particles and their interactions. However, it still fails to address several puzzles, particularly those revealed by astronomical observations. Among them, the nature of dark matter remains one of the most profound mysteries.


To date, all evidence for the existence of dark matter comes from its gravitational effects and the only precise quantitative measurement is its energy density, $\Omega_c \approx 0.26$, derived from its impact on the evolution of the Universe \cite{Planck:2018vyg}. Notably, the baryon energy density is $\Omega_b \approx 0.05$, which leads to the intriguing ratio $\Omega_c / \Omega_b \approx 5$, known as the dark matter-baryon coincidence problem. This coincidence is particularly surprising when compared to other ratios among non-relativistic species, such as the proton-to-neutron density ratio, $\Omega_p / \Omega_n \approx 7$, which is ensured by the approximate $SU(2)$ isospin symmetry. Such a striking similarity may hint at a deeper connection between the dark and visible sectors and offer new insights into the microscopic nature of dark matter.


To address the coincidence problem, a well-known idea is Asymmetric Dark Matter (ADM) \cite{Spergel:1984re,Nussinov:1985xr,Gelmini:1986zz,Barr:1990ca,Barr:1991qn,Kaplan:1991ah,Gudnason:2006ug,Gudnason:2006yj,Kaplan:2009ag,Davoudiasl:2012uw,Petraki:2013wwa,Zurek:2013wia}, where a conserved $U(1)$ global symmetry is introduced to ensure comparable number densities between the two sectors, $n_D \approx n_B$. However, a complete solution also requires an additional mechanism to explain the similarity in masses, $m_D \approx m_B$, which is the primary focus of this work.

The baryon mass $m_B$ is known to be determined by the QCD confinement scale $\Lambda_{\rm QCD}$, so several attempts have explored analogous constructions in the dark sector. These studies introduce a dark QCD sector, where dark baryon serves as the dark matter candidate and acquires its mass set by a dark confinement scale $\Lambda_{\rm dQCD}$. Previous approaches have focused on relating the two confinement scales, $\Lambda_{\rm dQCD} \approx \Lambda_{\rm QCD}$, through various ideas, including mirror dark sectors \cite{Hodges:1993yb,Berezhiani:1995am,Foot:2003jt,An:2009vq,Farina:2015uea,GarciaGarcia:2015pnn,Lonsdale:2018xwd,Ibe:2019ena,Bodas:2024idn}, infrared fixed points \cite{Bai:2013xga,Newstead:2014jva,Ritter:2022opo,Ritter:2024sqv}, and dark unification \cite{Murgui:2021eqf,Chung:2024nnj}.\footnote{The coincidence can also be achieved without comparable masses if the production and masses are correlated \cite{Rosa:2022sym}, or if novel dynamics emerge to balance the two energy densities \cite{Brzeminski:2023wza,Banerjee:2024xhn}.}


In this study, we consider a new approach in which the dark matter mass is generated from the QCD vacuum, naturally satisfying the condition $m_D \approx \Lambda_{\rm QCD}$. However, the measured running of QCD excludes the existence of new colored particles below $\mathcal{O}(100)$ GeV. Therefore, a dark matter candidate with a mass comparable to that of baryons must be colorless, raising an interesting question: \textit{Can a colorless particle still acquire a mass of $\mathcal{O}(\Lambda_{\rm QCD})$ from the QCD vacuum?} 

In the following, we systematically introduce the mechanisms required to realize this idea, culminating in a strongly coupled chiral dark sector with QCD-induced vacuum misalignment, which demonstrates how a colorless particle can still strongly couple to the QCD vacuum through a distinctive strongly coupled theory.

\section{$m_\nu$ from QCD in Pati-Salam model}

To illustrate how the QCD vacuum can generate masses for colorless particles, we begin by considering the example of neutrinos \cite{Thomas:1992hf,McDonald:1996cs,Ibanez:2001nd,Davoudiasl:2005ai,Babic:2019zqu,Davoudiasl:2021nfv}. By extending the SM gauge group $SU(3)_C\times U(1)_Y$ to the 
Pati-Salam subgroup $SU(4)_{PS}\times U(1)_R$ \cite{Pati:1974yy}, which features quark-lepton unification, neutrinos can couple to up-type quarks through
\begin{align}\label{PSint}
{\cal L}_{\text{E}}=\frac{1}{\sqrt{2}} g_{E} {E}_{\mu} (\bar{u}_L\gamma^\mu \nu_L+\bar{u}_R\gamma^\mu \nu_R)\,~,
\end{align}
where $g_E$ is the $SU(4)_{PS}$ gauge coupling and $E_\mu$ denotes the leptoquark field. Once the gauge group is broken at the scale $f_E$, the heavy leptoquark can be integrated out, generating effective four-fermion operators
\begin{align}
\mathcal{L}_{\text{E,eff}}\supset\frac{g_{E}^2}{M_{E}^2}
\left({\bar{\nu}_R}{\nu}_{L}\right)
\left({\bar{u}_L}{u}_{R}\right)+\cdots \text{(after Fierzing)}~
\end{align}
with the leptoquark mass $M_E = g_E f_E$.

Once QCD becomes strongly coupled and quarks form a condensate, a neutrino mass is dynamically generated. By naive dimensional analysis \cite{Weinberg:1978kz,Manohar:1983md,Georgi:1985hf}, the condensate can be estimated as
$\left<u\bar{u}\right> \approx  {\Lambda_{\rm QCD}^3}/{16\pi^2} \approx 4\pi\, v_{\rm QCD}^3~,$ where $v_{\rm QCD}$ is the chiral symmetry breaking scale of QCD, equivalent to $f_\pi$. The mass then takes the form
\begin{align}\label{deltamnu}
m_\nu \approx  \frac{g_{E}^2}{M_{E}^2} \left<u\bar{u}\right>\approx \frac{4\pi\, v_{\rm QCD}^3}{f_E^2}=4\pi
\left(\frac{v_{\rm QCD}}{f_E}\right)^2v_{\rm QCD}~,
\end{align}
demonstrating its origin from the QCD vacuum.

The prefactor in front of $v_{\rm QCD}$ can be interpreted as an effective Yukawa coupling between the neutrinos and the $v_{\rm QCD}$, defined as $Y_\nu \equiv 4\pi\,(v_{\rm QCD}/f_E)^2$. This coupling, however, is suppressed by the ratio of the QCD scale to the Pati–Salam scale. To construct a viable dark matter model, one must go beyond this setup by introducing new colored fermions $\psi$, whose condensate takes the form
\begin{align}\label{psipsi}
\langle \psi\bar{\psi} \rangle \approx 4\pi\,f_E^2\,v_{\rm QCD}~.
\end{align}
This condensate remains proportional to $v_{\rm QCD}$ but carries a prefactor of $4\pi f_E^2$, thereby avoiding the suppression. This form naturally leads us to consider the mechanism of \textbf{Vacuum Misalignment}.

\section{Toy model with QCD-induced vacuum misalignment}

Such a condensate features two scales, indicating the presence of a new strong dynamics in the dark sector, associated with a new dark color gauge group $SU(N)_D$ and a confinement scale $\Lambda_D \approx 4\pi f_D$, where $f_D$ denotes the symmetry breaking scale. Moreover, this structure is reminiscent of the condensate in composite Higgs models (CHMs) \cite{Kaplan:1983fs,Kaplan:1983sm}, which were proposed to address the electroweak (EW) hierarchy problem.

In CHMs (see \cite{Bellazzini:2014yua,Panico:2015jxa,Cacciapaglia:2014uja,Cacciapaglia:2020kgq} for reviews), the Higgs boson emerges as a pseudo-Nambu-Goldstone boson (pNGB) associated with the spontaneous breaking of an approximate global symmetry in a new strong sector based on a hypercolor gauge group with hyperfermions $Q$. Once the hypercolor group becomes strongly coupled, the condensate $\langle Q\bar{Q}\rangle  \approx 4\pi\,f_H^3$ forms, where $f_H$ is the symmetry breaking scale of the hypercolor sector. The condensate contains two distinct components: an EW-conserving one and an EW-violating one. The latter is proportional to the observed Higgs VEV, $v_{\rm EW} \approx 246~\text{GeV}$, and the SM fermions obtain their masses exclusively from this component, given by
\begin{align}\label{QQ}
\left<Q\bar{Q}\right>_{\cancel{EW}}  \approx 4\pi\,f_H^3\, {\rm sin}\,\theta~=4\pi\,f_H^2\, v_{\rm EW}~,
\end{align}
which is suppressed by a small angle $\theta$ due to vacuum misalignment, yielding a form similar to Eq.~\eqref{psipsi} \cite{Cacciapaglia:2015yra, Chung:2023iwj}. Furthermore, since SM fermions are hypercolorless particles that nevertheless acquire masses from the hypercolor vacuum in CHMs, this mechanism closely parallels the scenario we aim to realize in the dark sector.

To reproduce a similar condensate in the dark sector, we construct a model analogous to CHMs. The essential idea is to have a dark-fermion condensate formed at the scale $\Lambda_D\gg\Lambda_{\rm QCD}$ with a nearly flat direction associated with an approximate global symmetry, which is subsequently lifted by QCD effects at the scale $\Lambda_{\rm QCD}$, inducing a small vacuum misalignment.

Following this idea, we first introduce two Dirac dark quarks, $\Psi = (\psi, \psi')$, in our toy model. They are charged under both dark color $SU(N)_D$ and SM color $SU(3)_C$, transforming as $(N, 3)$. The dark color becomes strongly coupled first and confines at the scale $\Lambda_D$, leading to the condensate $\langle \Psi\bar{\Psi} \rangle \approx 4\pi f_D^3$, which breaks the approximate global symmetry $SU(6)_L \times SU(6)_R$ down to $SU(6)_V$, producing 35 pNGBs. Since the $SU(3)_C$ interaction explicitly breaks the global symmetry, 32 colored pNGBs, including two real color octets and one complex color octet, get a mass $\sim g_s\Lambda_D/4\pi$.

The rest is equivalent to the $SU(2)_L \times SU(2)_R / SU(2)_V$ coset with three dark pions, $\pi'_i$, corresponding to the three flat directions. The breaking can be parametrized by a non-linear sigma model, with $\Sigma$ transforming as a $(2, \bar{2})$ under $SU(2)_L \times SU(2)_R$. The most general condensate consistent with the coset is given by
\begin{equation}
\left<\Psi\bar{\Psi}\right>_{\rm D}\approx 4\pi f_D^2\Sigma\,,~\Sigma=f_D
\begin{pmatrix}
e^{i\alpha}{\rm cos}\,\theta  &  e^{i\beta}{\rm sin}\,\theta \\
-e^{i\beta}{\rm sin}\,\theta   &  e^{-i\alpha}{\rm cos}\,\theta \\
\end{pmatrix},
\end{equation}
where $\alpha,\beta,$ and $\theta$ parametrize the three flat directions.

\begin{figure}[tbp]
\centering
\includegraphics[width=0.45\textwidth]{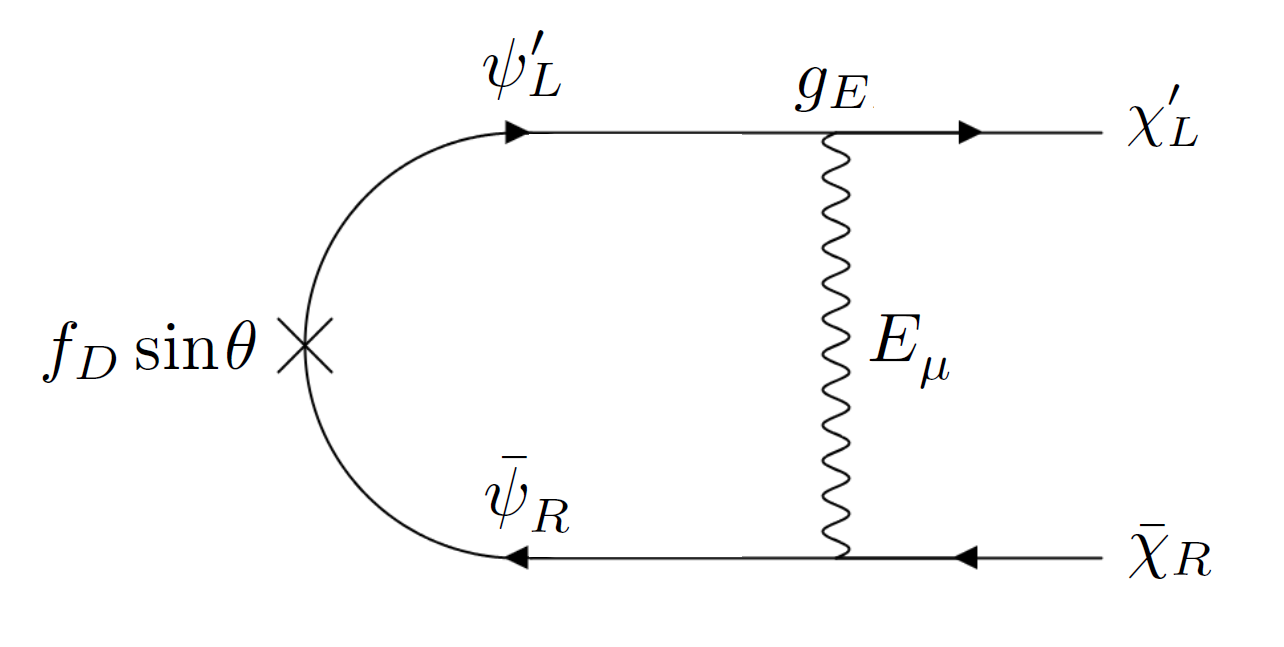}
\caption{The Feynman diagram responsible for the mass generation of the dark fermion $\chi$ in the toy model.\label{fig:DMloop}}
\end{figure}

Next, we introduce two dark Weyl fermions, $\chi'_L$ and $\chi_R$, which couples exclusively to specific dark quarks via 
\begin{align}\label{Ychi}
{\cal L}_{\text{E}}&=\frac{1}{\sqrt{2}} g_{E} {E}_{\mu} (\bar{\psi'}_L\gamma^\mu \chi'_L+\bar{\psi}_R\gamma^\mu \chi_R)~,
\end{align}
analogous to CHMs. $\chi'_L$ and $\chi_R$ can only acquire its mass from the $\langle \psi'_L \bar{\psi}_R \rangle$ part of the condensate, as illustrated in Fig.~\ref{fig:DMloop}. The resulting mass is given by
\begin{align}\label{chimass0}
m_\chi \approx  \frac{g_{E}^2}{M_{E}^2} \left<\psi'_L\bar{\psi}_R\right>=4\pi
\left(\frac{f_D}{f_E}\right)^2 f_D\, {\rm sin}\,\theta~.
\end{align}
A successful model with $m_\chi\approx \Lambda_{\rm QCD} \approx 4\pi\, v_{\rm QCD}$ requires the following two key conditions to be satisfied:
\begin{align}\label{condition}
\boldsymbol{\textbf{(1) }\,f_D\, {\rm sin}\,\theta\approx v_{\rm QCD}}~
\quad\text{and}\quad~
\boldsymbol{\textbf{(2) }\,f_E\approx f_D}~.
\end{align}


The first condition implies the vacuum misalignment triggered by QCD, which requires analyzing the pNGB potential. To study this, we take $(\alpha,\beta,\theta)=0$ as the initial vacuum around which to expand, and the three dark pion fields, $\pi'_i$, can then be described by the
\begin{equation}\label{Goldstone}
U(x)\equiv e^{\frac{i\pi'_i(x) \sigma_i}{f_D}},~
\pi'_i \sigma_i=
\begin{pmatrix}
\pi'_3 & \pi'_1-i\pi'_2 \\  
\pi'_1+i\pi'_2 & -\pi'_3 \\  
\end{pmatrix}~,
\end{equation}
where $\sigma_i$ are the broken generators. We denote $\pi'_0 \equiv \pi'_3$ as the real scalar and $\Pi'\equiv(\pi'_1 \pm i\pi'_2)/\sqrt{2}$ as a complex scalar. Since $\pi'_0$ will be eaten by a gauge boson in the UV-complete model discussed later, we focus only on the potential $V(\Pi')$ in this section.

The desired potential $V(\Pi')$ must be sufficiently flat for QCD effects to influence the vacuum at the scale $\Lambda_{\rm QCD}$. Therefore, the coefficient of the quadratic term $\mu^2$ should be smaller than $\Lambda_{\rm QCD}^2$. In the toy model, the interaction in Eq.~\eqref{Ychi} already generates a negative quadratic term of $\mathcal{O}(\Lambda_D^2)$, analogous to the top-loop contribution in CHMs. However, additional positive quadratic contributions can arise from four-Fermi interactions, such as $\left(\bar{\psi}_L \psi_R\right)^2$ and $\left(\bar{\psi}'_L \psi'_R\right)^2$, compensating for the negative term. A detailed discussion of these contributions and the cancellation between them within the UV-complete model is provided in the Appendix. After the quadratic terms cancel, the resulting potential is left with a quartic term given by
\begin{equation}\label{quartic}
V(\Pi')|_{\rm UV} \sim \lambda f_D^4 \text{ sin}^4\left(\frac{|\Pi'|}{f_D}\right)
~\text{with}~|\Pi'|=\sqrt{{\pi'_1}^2+{\pi'_2}^2}~.
\end{equation}

Next, when QCD becomes strongly coupled at the GeV scale, the dark quarks undergo an additional condensate formation. Since the dark color has already confined at the higher scale $\Lambda_D$, occupying the diagonal entries, the new condensate induced by QCD can only form along the orthogonal direction\footnote{The underlying physics involves the interaction between the two strong dynamics, which are non-perturbative. In this study, a specific phase is assumed based on theoretical insight. However, the exact physics should rely on first-principles calculations, such as Lattice or Functional Renormalization Group.}, giving
\begin{equation}\label{spurion}
\left<\Psi\bar{\Psi}\right>_{\rm QCD}\approx 4\pi\,v_{\rm QCD}^3 \times i\sigma_2 \equiv \mathbb{V}_{\rm QCD}~,
\end{equation}
where we assume $CP$ is conserved and the condensate happens along the $CP$-even direction. It then induces a backreaction that affects the original condensate $\left<\Psi\bar{\Psi}\right>_{\rm D}$, which can be described by the spurion $\mathbb{V}_{\rm QCD}$. Since this backreaction is orthogonal to the initial vacuum $\Sigma$, it destabilizes the vacuum along the $\pi'_2$ direction by generating a linear term in the potential. With the quartic term in Eq.~\eqref{quartic}, the complete potential is given by
\begin{equation}\label{Vpion}
V(\Pi') \sim - \Lambda_{\rm QCD}^3f_D \text{ sin}\left(\frac{\pi'_2}{f_D}\right)+ \lambda f_D^4 \text{ sin}^4\left(\frac{|\Pi'|}{f_D}\right).
\end{equation}

The $\pi'_2$ acquires a non-zero VEV, indicating that the vacuum shifts along the $\pi'_2$ direction and is stabilized by the quartic term, resulting in a small misalignment
\begin{equation}\label{misalignment}
f_D\, {\rm sin}\,\theta =f_D \text{ sin}\left(\frac{\langle \pi'_2\rangle}{f_D}\right) \approx \frac{\Lambda_{\rm QCD}}{\sqrt[3]{\lambda}}\approx
\frac{4\pi}{\sqrt[3]{\lambda}}\,v_{\rm QCD}~.
\end{equation}
The prefactor is determined by $\lambda$, which originates from the strong dynamics and is expected to be $\lambda\sim 16\pi^2$, with further details discussed in the Appendix. We obtain $f_D\, {\rm sin}\,\theta \approx 2.3\times v_{\rm QCD}$ as desired. In the new vacuum, the shift symmetry is further broken and the dark pions become massive, with $m_{\pi'_2}\approx\sqrt{3}\sqrt[6]{\lambda}\,\Lambda_{\rm QCD}\approx 4$ GeV and $m_{\pi'_1}\approx\sqrt[6]{\lambda}\,\Lambda_{\rm QCD} \approx 2.3$ GeV. They both play an important role in cosmology and phenomenology.


\section{DM as composite fermion in strongly coupled chiral theory}

With the first condition in Eq.~\eqref{condition} satisfied, we still require \textbf{(2) $\boldsymbol{f_D \approx f_E}$} to reproduce the desired condensate. This condition implies that the breaking of the extended gauge group must be driven by the dark color dynamics. Furthermore, it indicates that the two dark Weyl fermions, $\chi'_L$ and $\chi_R$, must carry dark color charge while remaining massless after dark color confinement. These requirements naturally motivate the consideration of \textbf{strongly coupled chiral gauge theories}, which feature massless composite fermions.

\begin{table}[t]
\centering
\begin{tabular}{|c||c|c|c|c|c|}
\hline
     & $SU(5)$   & $SU(4)$   & $SU(3)$ &  $U(1)_X$ & $U(1)_{D'}$  \\ \hline
$\quad A\quad $ & $A$   & $1$    & $1$  &  $0$ & $1/5$\\ \hline
$\bar{F}$ & $\overline{\square}$ & $\overline{\square}$ & $1$ & $-1/2$ & $-3/20$  \\ \hline
$F$  & $\square$ & $1$ & $\square$ &  $2/3$  & $0$ \\ \hline
\end{tabular}
\caption{Particle content of the SU(5) chiral gauge theory with three extra Dirac pairs.\label{tab:SU(5)1}}
\end{table}

We utilize the classic example \cite{Dimopoulos:1980hn}: an $SU(5)$ chiral theory containing one antisymmetric tensor $A$ and one anti-fundamental $\bar{F}$. This theory undergoes confinement without global symmetry breaking, resulting in a massless composite fermion, $A\bar{F}\bar{F}$. To incorporate QCD, we extend this minimal setup by adding three pairs of $F$, $\bar{F}$. The $SU(3)_C$ gauge group will be embedded into the enlarged global symmetry, as discussed below. The fermion content is summarized in Table~\ref{tab:SU(5)1}.

Although the dynamics of the theory remains unclear, several proposals for its phase structure have been discussed in \cite{Raby:1979my,Bars:1981se,Appelquist:2000qg} and recently in \cite{Csaki:2021xhi,Bolognesi:2021jzs,Smith:2021vbf,Karasik:2022gve,Li:2025tvu}. In this study, we adhere to a preserving $SU(5)$ gauge group but assume the $F\bar{F}$ condensate, which is a singlet under $SU(5)$, can still form. The scenario, discussed in \cite{Appelquist:2000qg}, features the symmetry breaking $SU(4) \times SU(3) \times U(1)_X \times U(1)_{D'} 
\to  SU(3)_V \times U(1)^2$ where $U(1)^2\subset SU(4)\times U(1)_X \times U(1)_{D'}$. Although the global symmetry is now broken, a massless composite fermion remains in the low-energy spectrum, accompanied by 15 Goldstone bosons.

Next, we gauge the entire global symmetry except for $U(1)_{D'}$, leading to the gauge symmetry breaking pattern:
\begin{align}\label{SB}
SU(4) \times SU(3)  \times U(1)_X 
\to  SU(3)_C \times U(1)_Y ~.
\end{align}
The 15 Goldstone bosons are all eaten, giving massive leptoquarks $E_\mu$, colorons $G'_\mu$, and a $Z'_\mu$. There are two unbroken $U(1)$'s. One is the hypercharge $U(1)_Y$ with $Y= X+1/6~T_{15}$, where we define $T_{15}=\text{diag}(-3,1,1,1)$. The other is the $U(1)_D$ with $D=D'+3/20~T_{15}$, which serves as the dark baryon number.

The introduction of additional gauge symmetries raises anomaly issues. Moreover, the current model contains a massless Weyl fermion and a colored Dirac fermion, which account for only half of the required fermions. To build a complete model, we introduce a duplicate sector, denoted as $SU(5)'$. Based on its gauge structure, we refer to the full framework as the “double 54321” model with the extended particle content summarized in Table \ref{tab:SU(5)2}.

The new $SU(5)'$ sector has the same fermion content as the original $SU(5)$ sector but with opposite charges. The two sectors subsequently undergo an additional symmetry breaking step, $SU(5)\times SU(5)'\to SU(5)_D$, to combine together. The resulting fermion content appears vector-like under $SU(5)_D$. However, as long as the four-fermion interactions mediated by the broken $SU(5)$ and $SU(5)'$ sectors are sufficiently strong, a chiral phase featuring massless composite fermions can still emerge.

\begin{table}[t]
\centering
\begin{tabular}{|c||c|c||c|c|c|c|c|}
\hline
& $SU(5)$   & $SU(5)'$  & $SU(5)_D$  & $SU(4)$   & $SU(3) $ & $U(1)_X$  & $U(1)_{D'}$\\ \hline
$ ~ A~ $       & $A$       & $1$     & $A$  & $1$    & $1$  & $0$ & $1/5$ \\ \hline
$\bar{F}$   & $\overline{\square}$ & $1$& $\overline{\square}$ & $\overline{\square}$  & $1$ & $-1/2$& $-3/20$\\ \hline
$F$         & $\square$ & $1$& $\square$ & $1$& $\square$  & $2/3$ & $0$ \\ \hline
$\bar{A}' $  & $1$ & $\bar{A}$   & $\bar{A}$& $1$    & $1$  & $0$ & $-1/5$\\ \hline
$F'$         & $1$ & $\square$& $\square$ & $\square$ & $1$  & $1/2$ & $3/20$\\ \hline
$\bar{F}'$         & $1$ & $\overline{\square}$ & $\overline{\square}$& $1$ & $\overline{\square}$  & $-2/3$ & $0$\\ \hline
\end{tabular}
\caption{Particle content of the double 54321 model.\label{tab:SU(5)2}}
\end{table}

The enlarged fermion content exhibits an enhanced global symmetry of $SU(7)_L \times SU(7)_R$ under $SU(5)_D$, which is subsequently broken to $SU(6)_V \times U(1)'$ after dark color confinement, including the gauge symmetry breaking in Eq.~\eqref{SB}. Because the $SU(4)$ is broken by the dark color dynamics, its breaking scale satisfies $f_E = f_D$, fulfilling the second condition in Eq.~\eqref{condition}. The symmetry breaking generates a total of 60 pNGBs. Among them, 15 are eaten by the heavy gauge bosons as their longitudinal components, including the previously mentioned $\pi'_0$. Another 42 carry QCD color charges and acquire masses of $\mathcal{O}(f_D)$. The remaining 3 neutral pNGBs present in the low-energy spectrum include a complex dark pion $\Pi'$ and a composite QCD axion $a\sim \left(\bar{F}_T\gamma^5F+\bar{F}'\gamma^5F'_T-6\,\bar{F}_S\gamma^5F_S\right)$ \cite{Kim:1984pt,Choi:1985cb,Kaplan:1985dv}, whose associated current is anomalous under QCD. We, therefore, obtain the axion interaction as
\begin{align}\label{axion}
\mathcal{L}_a = \frac{g_s^2}{32\pi^2}\frac{a}{f_a}G\tilde{G} ~~~\text{with}~~~ f_a=\frac{\sqrt{21}}{5}\,f_D~.
\end{align}

The fermion content after the symmetry breaking is summarized in Table \ref{tab:SU(5)3}, where we also indicate its correspondence with the toy model. The $F_T$ and $F_T'$, decomposed from $F$ and $F'$, have the desired quantum numbers required for $\psi$ and $\psi'$. The $F_S$ and $F_S'$ are identified with $\chi_R$ and $\chi'_L$. Together with the $E_\mu$ boson from the broken $SU(4)$ gauge symmetry that couples them as in Eq.~\eqref{Ychi}, we can reproduce the diagram in Fig.~\ref{fig:DMloop}, where the mass of $F_S$ is generated only through the part of the dark quark condensate $\langle {F'_T}\bar{F}_T\rangle$ that becomes nonzero due to the QCD-induced vacuum misalignment.

The dark matter is the composite fermion formed from $\bar{A}F_SF_S$, denoted by the dark baryon $X$, which is a gauge singlet but carries a dark baryon number of $D = 1$. Containing two $F_S$ fields, the mass is given by
\begin{align}\label{chimass}
m_X\approx 2\,m_{F_S} \approx  2\,\frac{g_{E}^2}{M_{E}^2} \left<{F'_T}\bar{F}_T\right> \approx
\frac{8\pi}{\sqrt[3]{\lambda}}\,\Lambda_{\rm QCD}~,
\end{align}
which is $\approx 4.6$ GeV for the benchmark choice. Comparable dark matter and baryon masses can then be explained by the fact that dark matter derives its mass from the QCD vacuum as baryons do.

\begin{table}[t]
\centering
\begin{tabular}{|c||c|c|c|c|c|}
\hline
&   $SU(5)_D$  & $SU(3)_C$  & $U(1)_Y$ &  $U(1)_D$ \\ \hline
$ \qquad A\qquad $       &  $A$ & $1$ & $0$ & $1/5$   \\ \hline
\tabincell{c}{$\bar{F}_S~(\bar{\chi}_R$)\\$\bar{F}_T~(\bar{\psi}_R)$}   & $\overline{\square}$ &\tabincell{c}{$1$\\$\overline{\square}$}  & \tabincell{c}{$0$\\$-2/3$} &  \tabincell{c}{$-3/5$\\$0$} \\ \hline
$F~(\psi_L)$         & $\square$ & $\square$  & $2/3$  & $0$ \\ \hline
$\bar{X}_R=A\bar{F}_S\bar{F}_S$ &  $1$ & $1$ & $0$  & $-1$   \\ \hline

$\bar{A}' $  & $\bar{A}'$ & $1$ & $0$  & $-1/5$ \\ \hline
\tabincell{c}{${F}'_S~ ({\chi}'_L$)\\${F}'_T~(\psi'_L)$}         &  $\square$ &\tabincell{c}{$1$\\${\square}$}   & \tabincell{c}{$0$\\$2/3$}  & \tabincell{c}{$3/5$\\$0$} \\ \hline
$\bar{F}'~(\bar{\psi}'_R)$         & $\overline{\square}$ & $\overline{\square}$ & $-2/3$  & $0$\\ \hline
${X}'_L=\bar{A}'F'_SF'_S$ &  $1$ & $1$ & $0$  & $1$   \\ \hline
\end{tabular}
\caption{Particle content of the double 54321 model after the symmetry breaking. The $\bar{F}$ and $F'$ are decomposed into $(\bar{F}_S,\bar{F}_T)$ and $(F'_S,F'_T)$, where $S$ and $T$ denote singlet and triplet representations under the unbroken $SU(3)_C$. \label{tab:SU(5)3}}
\end{table}

\section{Constraints}

Last, we examine the constraints from cosmology and phenomenology using our benchmark values, which can place bounds on the dark QCD scale $f_D$.

\subsection{Cosmology}

The dark pions are the lightest particles in our dark sector. If they are stable, they could overclose the Universe. Therefore, they must decay into SM particles before Big Bang Nucleosynthesis (BBN) in order to remain consistent with cosmological observations. In our model, the dark pions couple to SM gluons and photons through triangle loop diagrams involving the charged fermions $F_T$ and $F'_T$. Depending on their CP properties, the effective interactions are given by (we focus on the gluon coupling, which dominates)
\begin{align}\label{pion}
\mathcal{L}_{\pi'GG} = c_{GG}\frac{g_s^2}{32\pi^2}
\left(\frac{\pi'_2}{f_D}G{G}+\frac{\pi'_1}{f_D}G\tilde{G}\right)~,
\end{align}
where $c_{GG}$ is an $\mathcal{O}(1)$ coefficient, which we simply take to be unity. The lighter dark pion, $\pi'_1$, is the phenomenologically most relevant state. Its interaction is analogous to that of a gluonic axion-like particle (ALP), with the coupling strength determined by the scale $f_D$.

Through the interaction above, the entropy of the dark sector can be transferred to the visible sector via dark pion decays. For dark pions with $m_{\pi'}\gtrsim 2$ GeV, the hadronic decay modes dominate, including $\eta\pi\pi$, $KK\pi$, $\rho\rho$, $\omega\omega$, and others \cite{Aloni:2018vki}. Although the decay rates are subject to sizable uncertainties due to the complexity of hadronic resonance calculations, the dark pion lifetime can be approximately estimated as \cite{Bauer:2017ris,Bauer:2021mvw}
\begin{align}\label{pion_lifetime}
\tau_{\pi'}
\sim \left(\frac{\alpha_s^2}{8\pi^3}\frac{m_{\pi'}^{3}}{f_D^2}\right)^{-1}
= 2 \,s\left(\frac{f_D}{10^{11} \text{ GeV}}\right)^2\left(\frac{2 \text{ GeV}}{m_{\pi'}}\right)^3~,
\end{align}
where $\alpha_s=0.3$ at $2$ GeV is used. The constraint from BBN, requiring $\tau_{\pi'} \lesssim 1\,s$, thus sets an upper bound on the dark QCD scale, $f_D \lesssim 7\times10^{10}$ GeV.

\subsection{Phenomenology}

Three light degrees of freedom play important roles in the phenomenology, including the dark baryon dark matter $X$, the dark pion $\pi'$, and the QCD axion $a$. They place lower bounds on the scale $f_D$, which, when combined with the BBN upper bound derived above, leaves only a finite allowed parameter space.
~\\

\textbf{(1) QCD axion $\boldsymbol{a}$}: The most stringent bound comes from stellar evolution in globular clusters, which constrains the axion decay constant and, consequently, the scale $f_D\gtrsim 5\times 10^7$ GeV \cite{Ayala:2014pea,Dolan:2022kul,Caputo:2024oqc}. Combined with the upper bound from the BBN constraint, this leaves roughly three orders of magnitude of viable parameter space.

While the indirect bounds from astrophysical observations are stringent, direct searches for the QCD axion remain challenging. Since axion does not constitute the dark matter in our model, conventional haloscope searches are not applicable here. Instead, the relevant probes are helioscopes, which search for solar axions. The current limit from the CAST collaboration \cite{CAST:2017uph} is slightly weaker than the globular cluster bound but future large-scale helioscopes, such as IAXO \cite{IAXO:2019mpb}, are expected to surpass the astrophysical bound and probe QCD axions with $f_D$ up by an order of magnitude.
~\\

\textbf{(2) Dark pion $\boldsymbol{\pi'}$}: The dark pion in our model is equivalent to an $\mathcal{O}(1)$-GeV gluonic ALP, which has been extensively studied in \cite{Aloni:2018vki}. Current searches based on $B$-meson decays at LHCb \cite{LHCb:2015nkv,LHCb:2016awg} constrain the scale $f_D$ only up to $\sim 10^3~\text{GeV}$, which is far below the lower bound derived from the QCD axion.

However, the sensitivity can be significantly improved at future proton beam-dump experiments, such as SHiP \cite{SHiP:2015vad}. For dark pions with $m_{\pi'}\gtrsim 2$ GeV, the conventional production channels through proton bremsstrahlung and light meson decays are suppressed. Nevertheless, dark pions can still be efficiently produced via the gluon fusion process, allowing it to probe scales up to $f_D \sim 10^8~\text{GeV}$ \cite{Ovchynnikov:2025gpx}. Even greater sensitivities can be achieved by proposed future long-lived particle detectors at the LHC, such as MATHUSLA \cite{Curtin:2018mvb}.
~\\

\textbf{(3) Dark baryon dark matter $\boldsymbol{X}$}: In our model, the dark matter is composed of the dark baryon $X$. Dark matter searches can be performed through direct detection, since the dark baryon can scatter off SM nucleons through the exchange of dark pions, especially the CP-even dark pion $\pi'_2$. The interaction term between the dark baryon and the dark pion is given by
\begin{align}\label{dark_yukawa}
\mathcal{L}_{\pi'XX} = g_X\bar{X}\left(\pi'_2+i\gamma_5\pi'_1\right)X~,
\end{align}
where $g_X$ is the Yukawa coupling originated from the underlying strong dynamics. Its value should be determined using non-perturbative methods, which is beyond the scope of the present work. Therefore, we adopt the naive dimensional analysis estimate and take $g_X=4\pi$ for our numerical estimates.

Together with the interaction in Eq.~\eqref{pion}, we can then derive the dark matter–nucleon scattering cross section
\begin{align}\label{sigma}
\sigma_n =
\frac{0.19}{\pi}\,\left(\frac{m_n m_X}{m_n+m_X}\right)^2\,m_n^2\,\left(g_X\cdot\frac{1}{m_{\pi'_2}^2}\cdot\frac{1}{f_D}\right)^2
\end{align}
where $m_n$ is the nucleon mass. The resulting scattering cross section is then given by (we take $m_X=5$ GeV)
\begin{align}\label{sigmaN}
\sigma_n =8\times 10^{-46}~{\rm cm^2}\cdot
\left(\frac{g_X}{4\pi}\right)^2
\left(\frac{4~\text{GeV}}{m'_{\pi_2}}\right)^4\left(\frac{10^8~\text{GeV}}{f_D}\right)^2.
\end{align}
We can see that, once the constraint on $f_D$ from the QCD axion is taken into account, the predicted cross section is at the level of $\sim 10^{-45}\,\mathrm{cm}^2$, approaching the neutrino fog for $\mathcal{O}(5)$ GeV dark matter \cite{OHare:2021utq}. This makes direct detection particularly challenging. Nevertheless, we remain optimistic that future experimental breakthroughs may eventually overcome the neutrino fog.

On the other hand, indirect detection of dark matter is unlikely in our case because solving the coincidence problem requires the dark matter to be asymmetric, i.e. its relic abundance is determined by the asymmetry between $X$ and $\bar{X}$, analogous to that of the SM baryons, leading to negligible indirect detection signals.


\section{Conclusion}


In this study, we address the comparable-mass aspect of the coincidence problem with a new mechanism. Using a strongly coupled chiral dark sector with QCD-induced vacuum misalignment, we demonstrate how a colorless dark matter candidate can acquire an $\mathcal{O}(\Lambda_{\rm QCD})$ mass from the QCD vacuum. Our approach establishes a deep connection between the dark sector and the QCD sector through the rich structure of strong dynamics.


The model introduces an $\mathcal{O}(1)$ GeV dark baryon dark matter, a typical feature of solutions to the coincidence problem, accompanied by dark pions also at the $\mathcal{O}(1)$ GeV scale and a composite QCD axion at the meV scale. A distinguishing feature of the model is that all relevant couplings are governed by a single characteristic scale $f_D$ associated with the dark color dynamics. We find that a finite viable parameter space still exists.

Practically, the idea is difficult to probe directly because of the high scale of $f_D$, but it can be tested through searches for three light particles: the dark matter, the dark pion, and the QCD axion, at future experiments. If both the dark baryon dark matter and the dark pion are discovered at the $\mathcal{O}(1)$ GeV scale, comparable to the QCD baryon mass, together with the dark matter-baryon abundance coincidence, the three coincidences involving dark matter and QCD baryons could serve as strong hints to a QCD-related dark sector. Furthermore, if the QCD axion is also discovered with an interaction scale comparable to that inferred from dark matter direct detection and dark pion searches, this additional interaction-scale coincidence would provide an indication for the UV framework presented in this work.


In this study, we focus exclusively on the mass aspect of the coincidence problem, while a complete solution would require embedding the model within an ADM framework. Such a UV model typically introduces a mechanism for darko-baryo-genesis featuring a $U(1)$ symmetry that ensures comparable asymmetries are generated in both sectors. We leave the detailed construction of such a UV completion for future work but point out that dark grand unified theories \cite{Barr:2011cz,Barr:2013tea,Barr:2015lya,Chung:2024nnj} provide a promising starting point, as they naturally contain the ingredients required for the ADM mechanism, such as the $U(1)_{D-(B-L)}$ symmetry. Moreover, such extensions introduce bifundamental fermions similar to $F_T$, which play an essential role in our mechanism. More importantly, since the SM is chiral, extensions of this kind naturally give rise to a chiral dark sector, matching the structure required in our study. It would be highly interesting to investigate whether such a construction can be realized.


\section*{Acknowledgments}

I thank \'Alvaro Pastor-Guti\'errez, Hitoshi Murayama, Hsin-Chia Cheng, and Giacomo Cacciapaglia for useful discussions on the different issues of strong dynamics.

\appendix

\section{The double 54321 model as a UV completion}\label{sec:UV}

\begin{table*}[t]
\centering
\begin{tabular}{|c||c|c||c|c|c|c|c||c|c|c|}
\hline
& $SU(5)$   & $SU(5)'$  & $SU(5)_D$  & $SU(4)$   & $SU(3)$ & $U(1)_X$   & $U(1)'_{D}$& $SU(3)_C$ & $U(1)_Y$   & $U(1)_{D}$\\ \hline
$ ~A~ $       & $A$       & $1$     & $A$  & $1$    & $1$  & $0$  & $1/5$  & $1$  & $0$  & $1/5$\\ \hline
$\bar{F}=\begin{pmatrix} \bar{F}_S \\ \bar{F}_T    \end{pmatrix}$   & $\overline{\square}$ & $1$& $\overline{\square}$ & $\overline{\square}$  & $1$ & $-1/2$ & $-3/20$ &\tabincell{c}{$1$\\$\overline{\square}$}  & \tabincell{c}{$0$\\$-2/3$}&  \tabincell{c}{$-3/5$\\$0$}\\ \hline
$F$         & $\square$ & $1$& $\square$ & $1$& $\square$  & $2/3$  & $0$ & $\square$  & $2/3$  & $0$\\ \hline
$\bar{A}' $  & $1$ & $\bar{A}$   & $\bar{A}$& $1$    & $1$   & $0$ & $-1/5$ & $1$  & $0$  & $-1/5$\\ \hline
$F'=\begin{pmatrix} F'_S \\ F'_T    \end{pmatrix}$         & $1$ & $\square$& $\square$ & $\square$ & $1$  & $1/2$ & $3/20$&\tabincell{c}{$1$\\${\square}$}  & \tabincell{c}{$0$\\$2/3$} & \tabincell{c}{$3/5$\\$0$}\\ \hline
$\bar{F}'$  & $1$ & $\overline{\square}$ & $\overline{\square}$& $1$ & $\overline{\square}$  & $-2/3$ & $0$ & $\overline{\square}$  & $-2/3$ & $1/5$\\ \hline
\end{tabular}
\caption{Particle content of the double 54321 model with all the relevant quantum numbers listed.\label{tab:54321}}
\end{table*}

In this appendix, we present the detailed connection between the toy model and the UV model, namely the double 54321 model. We begin with the particle content of the double 54321 model. The initial gauge group is given by $SU(5) \times SU(5)' \times SU(4) \times SU(3) \times SU(2)_W \times U(1)_X$. Since none of the dark-sector particles carry $SU(2)_W$ charge, we omit this factor in the following discussion. The complete matter content, along with its quantum numbers under the gauge groups and the global $U(1)'_D$ and $U(1)_D$ symmetries, is summarized in Table \ref{tab:54321}.

Starting at a high scale where the vacuum $V_5$ emerges with $\langle V_5 \rangle = f_5$, this vacuum triggers the first gauge symmetry breaking $SU(5) \times SU(5)' \to SU(5)_D$, where $SU(5)_D$ is the dark color group of the theory. Next, $SU(5)_D$ becomes strongly coupled at the confinement scale $\Lambda_D \approx 4\pi f_D$, leading to the condensate of the form
\begin{align}
\left<\Psi_L\bar{\Psi}_R\right>\approx 4\pi\,f_D^3~,
\end{align}
where $\Psi_L=(F,F')=(F,F'_T,F'_S)$ and $\bar{\Psi}_R=(\bar{F}',\bar{F})=(\bar{F}',\bar{F}_T,\bar{F}_S)$. In this UV-complete model, the global symmetry associated with the $SU(5)_D$ dark color is $SU(7)_L \times SU(7)_R$, and there are two vacua of particular interest. Parameterizing the breaking through a non-linear sigma model, with $\Sigma$ transforming as a $(7,\bar{7})$ under $SU(7)_L \times SU(7)_R$, we can write these vacua in matrix form as follows
\begin{align}
&\Sigma_c = 
\bordermatrix{~ & ~ & \bar{F}' & ~ & ~ & \bar{F}_T & ~ & \bar{F}_S \cr
              ~ & 0 & 0 & 0 & 1 & 0 & 0 & 0\cr
              F & 0 & 0 & 0 & 0 & 1 & 0 & 0\cr
              ~ & 0 & 0 & 0 & 0 & 0 & 1 & 0\cr
              ~ & -1 & 0 & 0 & 0 & 0 & 0 & 0\cr
           F'_T & 0 & -1 & 0 & 0 & 0 & 0 & 0\cr
              ~ & 0 & 0 & -1 & 0 & 0 & 0 & 0\cr
           F'_S & 0 & 0 & 0 & 0 & 0 & 0 & 0\cr},\\
&\text{and }\Sigma_s = 
\bordermatrix{~ & ~ & \bar{F}' & ~ & ~ & \bar{F}_T & ~ & \bar{F}_S \cr
              ~ & 1 & 0 & 0 & 0 & 0 & 0 & 0\cr
              F & 0 & 1 & 0 & 0 & 0 & 0 & 0\cr
              ~ & 0 & 0 & 1 & 0 & 0 & 0 & 0\cr
              ~ & 0 & 0 & 0 & 1 & 0 & 0 & 0\cr
           F'_T & 0 & 0 & 0 & 0 & 1 & 0 & 0\cr
              ~ & 0 & 0 & 0 & 0 & 0 & 1 & 0\cr
           F'_S & 0 & 0 & 0 & 0 & 0 & 0 & 1\cr}~,
\end{align}
where we label the corresponding fermions on the sides of the matrix. The two vacua $\Sigma_c$ and $\Sigma_s$ correspond to the $\cos\theta$ vacuum and $\sin\theta$ vacuum as already shown in Eq.~(6) in the main text, with phases ignored here. Note that we express them in a basis different from that of Eq.~(6) to better reflect their structure in this UV model.

The first vacuum, the desired vacuum $\Sigma_c$, corresponds to a \textbf{chiral-dominant phase}, in which the theory behaves as a chiral theory with massless composite fermions. Although this vacuum may appear unnatural at first glance, it can be understood as arising from the combined dark-quark condensates $\langle F\bar{F}_T\rangle$ and $\langle F'_T\bar{F}'\rangle$ generated by the $SU(5)$ and $SU(5)'$ sectors. One therefore expects it to be the correct vacuum in the limit $f_5 \to f_D$. This vacuum breaks the global symmetry $SU(7)_L \times SU(7)_R$ down to $SU(6)_V \times U(1)'$, accompanied by the gauge symmetry breaking $SU(4) \times SU(3) \times U(1)_X \to SU(3)_C \times U(1)_Y$. However, the vacuum does not break the global symmetry maximally but instead leaves the dark fermions $\bar{F}_S$ and $F'_S$ massless, resulting in two massless composite Weyl fermions.

The $\Sigma_s$ vacuum, on the other hand, corresponds to a \textbf{vector-dominant phase}. In the limit where the effects of the two original $SU(5)$ sectors are neglected, for instance taking $f_5 \to \infty$, the theory under $SU(5)_D \times SU(4) \times SU(3) \times U(1)_X$ becomes exactly vector-like, and this vacuum is favored. In this phase, the theory undergoes a common chiral symmetry breaking $SU(7)_L \times SU(7)_R \to SU(7)_V$, while leaving the gauge symmetry $SU(4) \times SU(3) \times U(1)_X$ unbroken. The vacuum $\Sigma_s$ is aligned with the $SU(5) \times SU(5)'$ breaking vacuum $V_5$ and is therefore disfavored by the corresponding interactions. In this vacuum, the dark fermions acquire a mass that is identical to the dark quarks' masses, since they still form a complete fourplet of $SU(4)$. This property is directly related to the condition $f_E = f_D$, which is important for a successful dark matter model to realize our idea.

\section{The pNGB potential}

With two distinct vacua, we seek to determine which one is favored. To analyze the vacuum structure and misalignment, we begin from the $\Sigma_c$ vacuum and study the pNGB potential, examining whether it is stable or tachyonic. Under the $\Sigma_c$ vacuum, the global symmetry breaking $SU(7)_L \times SU(7)_R \to SU(6)_V \times U(1)'$ introduces a total of 60 pNGBs, corresponding to the broken generators $X_a$. The broken generators, together with the corresponding pNGB fields, are organized in matrix form as follows
\begin{align}
&i\pi_aX_a \!\cdot\! \Sigma_c\!=\nonumber\\
&\begin{pmatrix}
-\Pi'\cdot\mathbb{I}_{3\times 3} +{\Pi'_8}_a\cdot T_a & \pi'_0\cdot\mathbb{I}_{3\times 3} +{\pi'_8}_a\cdot T_a &  \phi_1  \\
\pi'_0\cdot\mathbb{I}_{3\times 3}+\tilde{\pi'_8}_a\cdot T_a  & \Pi'^*\cdot\mathbb{I}_{3\times 3}+{\Pi'_8}_a\cdot T_a &    \phi_2 \\
\tilde{\phi}^*_1  & \tilde{\phi}^*_2  & a \\
\end{pmatrix}~.
\end{align}
with pNGB fields transforming under the unbroken $SU(3)_C \times U(1)_Y$ gauge group as
\begin{align}
&\Pi'_8:\left(8,0\right),~~\Pi':\left(1,0\right),~~\pi'_8,\tilde{\pi}'_8:\left(8,0\right),~~\pi'_0:\left(1,0\right),~~\nonumber\\
&\phi_1,\tilde{\phi}_1:\left(3,\frac{2}{3}\right),~~
\phi_2,\tilde{\phi}_2:\left(3,\frac{2}{3}\right),~~
a:(1,0)~.
\end{align}
The fields labeled by capital $\Pi$ are complex, and $\pi'_0$ and $\Pi'$ are the pNGBs of the $SU(2)_L \times SU(2)_R / SU(2)_V$ subcoset mentioned in the toy model. Among them, 15 are eaten by the gauge bosons of the broken gauge symmetry $SU(4) \times SU(3) \times U(1)_X  \to SU(3)_C \times U(1)_Y$, including a linear combination of $\phi_1$'s, a linear combination of $\pi'_8$'s, and the singlet $\pi'_0$. We are left with 45 pNGBs, along with massive leptoquarks $E_\mu$, colorons $G'_\mu$, and a $Z'_\mu$ boson.

Next, we discuss the pNGB potential arising from various sources of explicit global symmetry breaking, including gauge interactions and four-fermion operators induced by broken gauge symmetries. Owing to the nonlinear realization of the pNGB fields, the generic potential takes a trigonometric form as
\begin{equation}\label{potential}
V(\pi_a) = \mu_a^2 f_D^2 \text{ sin}^2\left(\frac{\pi_a}{f_D}\right)+ \lambda_a f_D^4 \text{ sin}^4\left(\frac{\pi_a}{f_D}\right)
~,
\end{equation}
where the first and second terms play roles analogous to the quadratic and quartic terms in a conventional scalar potential. In the following, we derive the coefficients $\mu_a^2$ and $\lambda_a$ for each pNGB from the corresponding contributions.

Starting with the gauge interaction from $SU(3)_C \times U(1)_Y$, the partially gauged symmetry explicitly breaks the global symmetry and thus contributes to the potential of the pNGB fields that are charged under it. For different gauge groups, the $\mu_a^2$ terms generated for the various pNGBs are given by the following
\begin{align}
SU(3)_C:~& \sim c_s\frac{9}{16\pi^2}g_s^2M_\rho^2 \text{ (for $\pi'_8,\Pi'_8$)}~,\nonumber\\
&\sim c_s\frac{1}{4\pi^2}g_s^2M_\rho^2 \text{ (for $\phi$'s)}~,\nonumber\\
U(1)_Y:~&\sim c_y\frac{1}{12\pi^2}g_Y^2M_\rho^2 \text{ (for $\phi$'s)}~,
\label{Vgauge}\end{align}
where the $c$'s are $\mathcal{O}(1)$ non-perturbative coefficients, $g_s$ and $g_Y$ are couplings of $SU(3)_C$ and $U(1)_Y$, and $M_\rho \sim g_\rho f_D$ is the mass of the composite resonances $\rho$, which the gauge bosons couple with. After 42 of the 45 pNGBs acquire masses of $\mathcal{O}(f_D)$ from gauge boson loops, three light pNGBs remain: the complex scalar $\Pi'$ and the QCD axion $a$. As the QCD axion potential arising from its anomalous coupling to gluons is well studied, we focus in the following on the potential $V(\Pi')$, denoting its coefficients by $\mu^2$ and $\lambda$.

Next, we move on to four-fermion interactions. These include the exchange of the massive leptoquarks $E_\mu$, the colorons $G'\mu$, the $Z'\mu$ boson, and the broken $SU(5) \times SU(5)'$. Since we now focus only on $V(\Pi')$, the effects of the coloron $G'\mu$ and $Z'\mu$ are rather trivial, as they can be treated as gauging broken subgroups of $SU(6)_L \times SU(6)_R$, leading to contributions similar to Eq.~\eqref{Vgauge}, given by
\begin{align}\label{Vbroken}
G'_\mu, Z'_\mu:~ \Delta\mu^2\sim 
-c_{G'}\frac{9}{16\pi^2}g_{G'}^2M_\rho^2
-c_{Z'}\frac{3}{16\pi^2}g_{Z'}^2M_\rho^2~.
\end{align}
The main difference compared to Eq.~\eqref{Vgauge} is that, since both interactions correspond to the broken gauge symmetry of the initial vacuum $\Sigma_c$, they give negative contributions, destabilizing the initial vacuum. Note that $\lambda$ is also generated but is suppressed as $\mathcal{O}(g^4/16\pi^2)$. We therefore retain only the dominant contribution that will be shown next.

The leptoquark $E_\mu$ interaction is rather tricky because it couples dark quarks $\bar{F}_T, F'_T$ to dark fermions $\bar{F}_S, F'_S$, which are outside the $SU(6)_L \times SU(6)_R \to SU(6)_V$ breaking pattern. In the main text, it is identified as the interaction in Eq.~(7), which couples $\chi'_L$ and $\chi_R$ to the condensate $\left<\psi'_L \bar{\psi}_R \right>$. The leptoquark interaction, under the $\Sigma_c$ vacuum, can be interpreted as the Yukawa coupling between $\bar{F}_S, F'S$ and the dark pion $\Pi'$. Therefore, its contribution to the pNGB potential is the same as that from the fermion loops of $\bar{F}S, F'S$ with a strong Yukawa coupling $Y_F \sim 4\pi \left( f_D^2 / f_E^2 \right) \sim 4\pi$, similar to the top quark loop contribution to the pNGB Higgs potential in CHMs. It thus gives a negative quadratic contribution and a large positive quartic term given by
\begin{align}\label{VYukawa}
E_\mu: \Delta\mu^2\sim 
-c_{F}\frac{Y_{F}^2}{16\pi^2}\Lambda_D^2
\sim -c_F \Lambda_D^2~,~
\lambda\sim \frac{Y_F^4}{16\pi^2} \sim 16\pi^2,
\end{align}
where $\Lambda_D$ is the cutoff of the fermion loops. This contribution is currently the dominant one due to the large Yukawa coupling $Y_F$. We also display the corresponding quartic term, which provides the leading contribution among all sources. This strong Yukawa interaction is a key ingredient of our mechanism, enabling successful dark matter mass generation from the QCD vacuum. At the same time, it is the strongest effect that disfavors the $\Sigma_c$ vacuum.

So far, most contributions are negative, making the $\Sigma_c$ vacuum unstable, which is no surprise since most gauge symmetries are broken in the $\Sigma_c$ vacuum. But, as we already mentioned, the dominant contribution should come from the four-fermion interactions arising from the broken $SU(5) \times SU(5)'$ gauge interaction that stabilizes the chiral-dominant vacuum $\Sigma_c$. With the breaking scale $f_5$, several four-fermion interactions are introduced as
\begin{align}
\mathcal{L}_{\rm 4-Fermi}= \frac{1}{f_5^2}&[\text{cos}^4\beta(\bar{F}_TF)^2+\text{sin}^4\beta(\bar{F'}F'_T)\nonumber\\
&+2\,\text{cos}^2\beta\,\text{sin}^2\beta(\bar{F}_TF)(\bar{F'}F'_T)]~, 
\end{align}
where tan$\beta={g'_5}/{g_5}$ These terms describe the strength of $SU(5) \times SU(5)'$, which pushes the true vacuum toward the $\Sigma_c$ vacuum. Their effect on the pNGB potential is also straightforward, i.e., they generate an additional positive quadratic term as
\begin{align}\label{VSU(5)}
SU(5)\times SU(5)':~ \Delta\mu^2\sim 
 ~c_5\frac{16\pi^2f^4_D}{f^2_5}\sim c_5\left(\frac{f^2_D}{f^2_5}\right)\Lambda_D^2~.
\end{align}
The contribution occurs at tree level without loop suppression, and is of the same order as Eq.~\eqref{VYukawa}. However, it is suppressed by a factor of $(f_D / f_5)^2$. To avoid this suppression, $f_5 \sim f_D$ is required, which implies that both $SU(5)$ and $SU(5)'$ couplings should become strong before the breaking. The condition can be met by introducing a $Z_2$ symmetry between the two sectors, with the breaking triggered dynamically by themselves \cite{Martin:1992mj,Arkani-Hamed:2000ulw}.

With all the contributions derived, we obtain the potential of the complex dark pion $V(\Pi')$ with coefficients given by
\begin{align}\label{Vpi}
\mu^2 \sim  
c_5\left(\frac{f^2_D}{f^2_5}\right)\Lambda_D^2-c_F \Lambda_D^2&
-c_{G'}\frac{3g_{G'}^2}{16\pi^2}M_\rho^2-c_{Z'}\frac{g_{Z'}^2}{16\pi^2}M_\rho^2,\nonumber\\
\lambda \sim& \frac{Y_F^4}{16\pi^2}\sim 16\pi^2~.
\end{align}
For the desired flat potential, the coefficient of the quadratic term must satisfy $\mu^2 < \Lambda_{\rm QCD}^2$. In the present realization, this requires a cancellation at the level of $\mathcal{O}(\Lambda_{\rm QCD}^2 / \Lambda_D^2)$, similar to the hierarchy problem of the scalar Higgs boson, or more precisely, the little hierarchy problem in composite Higgs models \cite{Kaplan:1983fs,Kaplan:1983sm}, where the SM Higgs also arises as a pNGB of a strongly coupled theory. However, the scalar potential could be rendered natural in a more sophisticated UV completion with enhanced symmetry.

As an illustration, by introducing two additional pairs of $F,\bar{F}$ for each $SU(5)$ sector, the global symmetry can be enlarged to $SU(6)\times SU(5)$, where we gauge the $SU(5)''\subset SU(6)$ subgroup. After introducing additional fermions $A''$ and $\bar{F}''$ charged under $SU(5)''$, one can impose a $Z_3$ symmetry among the three $SU(5)$ sectors. As a result, the corresponding vacua exhibit similar symmetry-breaking patterns and vacuum energies, leaving a flat direction associated with light pNGBs. More generally, mechanisms developed to address the electroweak hierarchy problem, such as collective symmetry breaking \cite{ArkaniHamed:2001nc,ArkaniHamed:2002qx,ArkaniHamed:2002qy}, discrete symmetries \cite{Hook:2018jle}, or other constructions \cite{Csaki:2017eio,Csaki:2019coc,Cheng:2020dum}, can also be adapted to this framework and help alleviate the tuning. We leave the detailed construction of a UV theory free from the hierarchy problem for future work.

In the final step, we derive the potential induced by the spurion $\mathbb{V}_{\rm QCD}$, which arises after QCD confinement. Its effect is analogous to that of a quark mass term, although the resulting consequences are quite different. The leading-order potential is given by
\begin{align}
~\Delta V(\Pi') \sim c'\Lambda_D\, {\rm tr}(\mathbb{V}_{\rm QCD}\,U)
\sim -\Lambda_{\rm QCD}^3f_D \text{ sin}\left(\frac{\pi'_2}{f_D}\right)~.
\end{align}
This linear term, corresponding to the direction associated with $\mathbb{V}_{\rm QCD}$, destabilize the vacuum along the $\pi'_2$ direction. The effect of the spurion can be understood as shifting the system from the original vacuum determined by the dark color dynamics to a new vacuum that is a combination of the dark color and QCD vacua. However, because $SU(2)_L \times SU(2)_R$ is already broken by the quartic interaction with coefficient $\lambda$, the true vacuum is stabilized at an intermediate point between these two configurations. Consequently, the dark pions acquire masses of $\sqrt{3}\sqrt[6]{\lambda}\,\Lambda_{\rm QCD}$ for CP-even $\pi'_2$ and $\sqrt[6]{\lambda}\,\Lambda_{\rm QCD}$ for CP-odd $\pi'_1$, respectively. For the phenomenological analysis in the main text, we focus exclusively on $\pi'_1$, which is the lightest dark pion.

Finally, for the composite QCD axion $a$, no potential term has been generated above the QCD confinement scale, reflecting that the corresponding symmetry is preserved by the gauge interactions. However, since $a\sim \left(\bar{F}_T\gamma^5F+\bar{F}'\gamma^5F'_T-6\,\bar{F}_S\gamma^5F_S\right)$ is associated with a current that is anomalous under QCD, non-perturbative QCD effects generate the well-known axion potential and mass given by
\begin{align}
&V(a)=-m_\pi^2f_\pi^2\sqrt{1-\frac{4m_um_d}{(m_u+m_d)^2}{\rm sin}^2\left(\frac{a}{2f_a}\right)}\nonumber\\
&\implies m_a^2 = \frac{4m_um_d}{(m_u+m_d)^2}\frac{m_\pi^2f_\pi^2}{f_a^2}\quad {\rm where}\quad f_a=\frac{\sqrt{21}}{5}\,f_D~. 
\end{align}

\bibliography{DM_Ref}

\begin{thebibliography}{87}%
\makeatletter
\providecommand \@ifxundefined [1]{%
 \@ifx{#1\undefined}
}%
\providecommand \@ifnum [1]{%
 \ifnum #1\expandafter \@firstoftwo
 \else \expandafter \@secondoftwo
 \fi
}%
\providecommand \@ifx [1]{%
 \ifx #1\expandafter \@firstoftwo
 \else \expandafter \@secondoftwo
 \fi
}%
\providecommand \natexlab [1]{#1}%
\providecommand \enquote  [1]{``#1''}%
\providecommand \bibnamefont  [1]{#1}%
\providecommand \bibfnamefont [1]{#1}%
\providecommand \citenamefont [1]{#1}%
\providecommand \href@noop [0]{\@secondoftwo}%
\providecommand \href [0]{\begingroup \@sanitize@url \@href}%
\providecommand \@href[1]{\@@startlink{#1}\@@href}%
\providecommand \@@href[1]{\endgroup#1\@@endlink}%
\providecommand \@sanitize@url [0]{\catcode `\\12\catcode `\$12\catcode `\&12\catcode `\#12\catcode `\^12\catcode `\_12\catcode `\%12\relax}%
\providecommand \@@startlink[1]{}%
\providecommand \@@endlink[0]{}%
\providecommand \url  [0]{\begingroup\@sanitize@url \@url }%
\providecommand \@url [1]{\endgroup\@href {#1}{\urlprefix }}%
\providecommand \urlprefix  [0]{URL }%
\providecommand \Eprint [0]{\href }%
\providecommand \doibase [0]{http://dx.doi.org/}%
\providecommand \selectlanguage [0]{\@gobble}%
\providecommand \bibinfo  [0]{\@secondoftwo}%
\providecommand \bibfield  [0]{\@secondoftwo}%
\providecommand \translation [1]{[#1]}%
\providecommand \BibitemOpen [0]{}%
\providecommand \bibitemStop [0]{}%
\providecommand \bibitemNoStop [0]{.\EOS\space}%
\providecommand \EOS [0]{\spacefactor3000\relax}%
\providecommand \BibitemShut  [1]{\csname bibitem#1\endcsname}%
\let\auto@bib@innerbib\@empty
\bibitem [{\citenamefont {Aghanim}\ \emph {et~al.}(2020)\citenamefont {Aghanim} \emph {et~al.}}]{Planck:2018vyg}%
  \BibitemOpen
  \bibfield  {author} {\bibinfo {author} {\bibfnamefont {N.}~\bibnamefont {Aghanim}} \emph {et~al.} (\bibinfo {collaboration} {Planck}),\ }\href {\doibase 10.1051/0004-6361/201833910} {\bibfield  {journal} {\bibinfo  {journal} {Astron. Astrophys.}\ }\textbf {\bibinfo {volume} {641}},\ \bibinfo {pages} {A6} (\bibinfo {year} {2020})},\ \bibinfo {note} {[Erratum: Astron.Astrophys. 652, C4 (2021)]},\ \Eprint {http://arxiv.org/abs/1807.06209} {arXiv:1807.06209 [astro-ph.CO]} \BibitemShut {NoStop}%
\bibitem [{\citenamefont {Spergel}\ and\ \citenamefont {Press}(1985)}]{Spergel:1984re}%
  \BibitemOpen
  \bibfield  {author} {\bibinfo {author} {\bibfnamefont {D.~N.}\ \bibnamefont {Spergel}}\ and\ \bibinfo {author} {\bibfnamefont {W.~H.}\ \bibnamefont {Press}},\ }\href {\doibase 10.1086/163336} {\bibfield  {journal} {\bibinfo  {journal} {Astrophys. J.}\ }\textbf {\bibinfo {volume} {294}},\ \bibinfo {pages} {663} (\bibinfo {year} {1985})}\BibitemShut {NoStop}%
\bibitem [{\citenamefont {Nussinov}(1985)}]{Nussinov:1985xr}%
  \BibitemOpen
  \bibfield  {author} {\bibinfo {author} {\bibfnamefont {S.}~\bibnamefont {Nussinov}},\ }\href {\doibase 10.1016/0370-2693(85)90689-6} {\bibfield  {journal} {\bibinfo  {journal} {Phys. Lett. B}\ }\textbf {\bibinfo {volume} {165}},\ \bibinfo {pages} {55} (\bibinfo {year} {1985})}\BibitemShut {NoStop}%
\bibitem [{\citenamefont {Gelmini}\ \emph {et~al.}(1987)\citenamefont {Gelmini}, \citenamefont {Hall},\ and\ \citenamefont {Lin}}]{Gelmini:1986zz}%
  \BibitemOpen
  \bibfield  {author} {\bibinfo {author} {\bibfnamefont {G.~B.}\ \bibnamefont {Gelmini}}, \bibinfo {author} {\bibfnamefont {L.~J.}\ \bibnamefont {Hall}}, \ and\ \bibinfo {author} {\bibfnamefont {M.~J.}\ \bibnamefont {Lin}},\ }\href {\doibase 10.1016/0550-3213(87)90424-X} {\bibfield  {journal} {\bibinfo  {journal} {Nucl. Phys. B}\ }\textbf {\bibinfo {volume} {281}},\ \bibinfo {pages} {726} (\bibinfo {year} {1987})}\BibitemShut {NoStop}%
\bibitem [{\citenamefont {Barr}\ \emph {et~al.}(1990)\citenamefont {Barr}, \citenamefont {Chivukula},\ and\ \citenamefont {Farhi}}]{Barr:1990ca}%
  \BibitemOpen
  \bibfield  {author} {\bibinfo {author} {\bibfnamefont {S.~M.}\ \bibnamefont {Barr}}, \bibinfo {author} {\bibfnamefont {R.~S.}\ \bibnamefont {Chivukula}}, \ and\ \bibinfo {author} {\bibfnamefont {E.}~\bibnamefont {Farhi}},\ }\href {\doibase 10.1016/0370-2693(90)91661-T} {\bibfield  {journal} {\bibinfo  {journal} {Phys. Lett. B}\ }\textbf {\bibinfo {volume} {241}},\ \bibinfo {pages} {387} (\bibinfo {year} {1990})}\BibitemShut {NoStop}%
\bibitem [{\citenamefont {Barr}(1991)}]{Barr:1991qn}%
  \BibitemOpen
  \bibfield  {author} {\bibinfo {author} {\bibfnamefont {S.~M.}\ \bibnamefont {Barr}},\ }\href {\doibase 10.1103/PhysRevD.44.3062} {\bibfield  {journal} {\bibinfo  {journal} {Phys. Rev. D}\ }\textbf {\bibinfo {volume} {44}},\ \bibinfo {pages} {3062} (\bibinfo {year} {1991})}\BibitemShut {NoStop}%
\bibitem [{\citenamefont {Kaplan}(1992)}]{Kaplan:1991ah}%
  \BibitemOpen
  \bibfield  {author} {\bibinfo {author} {\bibfnamefont {D.~B.}\ \bibnamefont {Kaplan}},\ }\href {\doibase 10.1103/PhysRevLett.68.741} {\bibfield  {journal} {\bibinfo  {journal} {Phys. Rev. Lett.}\ }\textbf {\bibinfo {volume} {68}},\ \bibinfo {pages} {741} (\bibinfo {year} {1992})}\BibitemShut {NoStop}%
\bibitem [{\citenamefont {Gudnason}\ \emph {et~al.}(2006{\natexlab{a}})\citenamefont {Gudnason}, \citenamefont {Kouvaris},\ and\ \citenamefont {Sannino}}]{Gudnason:2006ug}%
  \BibitemOpen
  \bibfield  {author} {\bibinfo {author} {\bibfnamefont {S.~B.}\ \bibnamefont {Gudnason}}, \bibinfo {author} {\bibfnamefont {C.}~\bibnamefont {Kouvaris}}, \ and\ \bibinfo {author} {\bibfnamefont {F.}~\bibnamefont {Sannino}},\ }\href {\doibase 10.1103/PhysRevD.73.115003} {\bibfield  {journal} {\bibinfo  {journal} {Phys. Rev. D}\ }\textbf {\bibinfo {volume} {73}},\ \bibinfo {pages} {115003} (\bibinfo {year} {2006}{\natexlab{a}})},\ \Eprint {http://arxiv.org/abs/hep-ph/0603014} {arXiv:hep-ph/0603014} \BibitemShut {NoStop}%
\bibitem [{\citenamefont {Gudnason}\ \emph {et~al.}(2006{\natexlab{b}})\citenamefont {Gudnason}, \citenamefont {Kouvaris},\ and\ \citenamefont {Sannino}}]{Gudnason:2006yj}%
  \BibitemOpen
  \bibfield  {author} {\bibinfo {author} {\bibfnamefont {S.~B.}\ \bibnamefont {Gudnason}}, \bibinfo {author} {\bibfnamefont {C.}~\bibnamefont {Kouvaris}}, \ and\ \bibinfo {author} {\bibfnamefont {F.}~\bibnamefont {Sannino}},\ }\href {\doibase 10.1103/PhysRevD.74.095008} {\bibfield  {journal} {\bibinfo  {journal} {Phys. Rev. D}\ }\textbf {\bibinfo {volume} {74}},\ \bibinfo {pages} {095008} (\bibinfo {year} {2006}{\natexlab{b}})},\ \Eprint {http://arxiv.org/abs/hep-ph/0608055} {arXiv:hep-ph/0608055} \BibitemShut {NoStop}%
\bibitem [{\citenamefont {Kaplan}\ \emph {et~al.}(2009)\citenamefont {Kaplan}, \citenamefont {Luty},\ and\ \citenamefont {Zurek}}]{Kaplan:2009ag}%
  \BibitemOpen
  \bibfield  {author} {\bibinfo {author} {\bibfnamefont {D.~E.}\ \bibnamefont {Kaplan}}, \bibinfo {author} {\bibfnamefont {M.~A.}\ \bibnamefont {Luty}}, \ and\ \bibinfo {author} {\bibfnamefont {K.~M.}\ \bibnamefont {Zurek}},\ }\href {\doibase 10.1103/PhysRevD.79.115016} {\bibfield  {journal} {\bibinfo  {journal} {Phys. Rev. D}\ }\textbf {\bibinfo {volume} {79}},\ \bibinfo {pages} {115016} (\bibinfo {year} {2009})},\ \Eprint {http://arxiv.org/abs/0901.4117} {arXiv:0901.4117 [hep-ph]} \BibitemShut {NoStop}%
\bibitem [{\citenamefont {Davoudiasl}\ and\ \citenamefont {Mohapatra}(2012)}]{Davoudiasl:2012uw}%
  \BibitemOpen
  \bibfield  {author} {\bibinfo {author} {\bibfnamefont {H.}~\bibnamefont {Davoudiasl}}\ and\ \bibinfo {author} {\bibfnamefont {R.~N.}\ \bibnamefont {Mohapatra}},\ }\href {\doibase 10.1088/1367-2630/14/9/095011} {\bibfield  {journal} {\bibinfo  {journal} {New J. Phys.}\ }\textbf {\bibinfo {volume} {14}},\ \bibinfo {pages} {095011} (\bibinfo {year} {2012})},\ \Eprint {http://arxiv.org/abs/1203.1247} {arXiv:1203.1247 [hep-ph]} \BibitemShut {NoStop}%
\bibitem [{\citenamefont {Petraki}\ and\ \citenamefont {Volkas}(2013)}]{Petraki:2013wwa}%
  \BibitemOpen
  \bibfield  {author} {\bibinfo {author} {\bibfnamefont {K.}~\bibnamefont {Petraki}}\ and\ \bibinfo {author} {\bibfnamefont {R.~R.}\ \bibnamefont {Volkas}},\ }\href {\doibase 10.1142/S0217751X13300287} {\bibfield  {journal} {\bibinfo  {journal} {Int. J. Mod. Phys. A}\ }\textbf {\bibinfo {volume} {28}},\ \bibinfo {pages} {1330028} (\bibinfo {year} {2013})},\ \Eprint {http://arxiv.org/abs/1305.4939} {arXiv:1305.4939 [hep-ph]} \BibitemShut {NoStop}%
\bibitem [{\citenamefont {Zurek}(2014)}]{Zurek:2013wia}%
  \BibitemOpen
  \bibfield  {author} {\bibinfo {author} {\bibfnamefont {K.~M.}\ \bibnamefont {Zurek}},\ }\href {\doibase 10.1016/j.physrep.2013.12.001} {\bibfield  {journal} {\bibinfo  {journal} {Phys. Rept.}\ }\textbf {\bibinfo {volume} {537}},\ \bibinfo {pages} {91} (\bibinfo {year} {2014})},\ \Eprint {http://arxiv.org/abs/1308.0338} {arXiv:1308.0338 [hep-ph]} \BibitemShut {NoStop}%
\bibitem [{\citenamefont {Hodges}(1993)}]{Hodges:1993yb}%
  \BibitemOpen
  \bibfield  {author} {\bibinfo {author} {\bibfnamefont {H.~M.}\ \bibnamefont {Hodges}},\ }\href {\doibase 10.1103/PhysRevD.47.456} {\bibfield  {journal} {\bibinfo  {journal} {Phys. Rev. D}\ }\textbf {\bibinfo {volume} {47}},\ \bibinfo {pages} {456} (\bibinfo {year} {1993})}\BibitemShut {NoStop}%
\bibitem [{\citenamefont {Berezhiani}\ \emph {et~al.}(1996)\citenamefont {Berezhiani}, \citenamefont {Dolgov},\ and\ \citenamefont {Mohapatra}}]{Berezhiani:1995am}%
  \BibitemOpen
  \bibfield  {author} {\bibinfo {author} {\bibfnamefont {Z.~G.}\ \bibnamefont {Berezhiani}}, \bibinfo {author} {\bibfnamefont {A.~D.}\ \bibnamefont {Dolgov}}, \ and\ \bibinfo {author} {\bibfnamefont {R.~N.}\ \bibnamefont {Mohapatra}},\ }\href {\doibase 10.1016/0370-2693(96)00219-5} {\bibfield  {journal} {\bibinfo  {journal} {Phys. Lett. B}\ }\textbf {\bibinfo {volume} {375}},\ \bibinfo {pages} {26} (\bibinfo {year} {1996})},\ \Eprint {http://arxiv.org/abs/hep-ph/9511221} {arXiv:hep-ph/9511221} \BibitemShut {NoStop}%
\bibitem [{\citenamefont {Foot}\ and\ \citenamefont {Volkas}(2003)}]{Foot:2003jt}%
  \BibitemOpen
  \bibfield  {author} {\bibinfo {author} {\bibfnamefont {R.}~\bibnamefont {Foot}}\ and\ \bibinfo {author} {\bibfnamefont {R.~R.}\ \bibnamefont {Volkas}},\ }\href {\doibase 10.1103/PhysRevD.68.021304} {\bibfield  {journal} {\bibinfo  {journal} {Phys. Rev. D}\ }\textbf {\bibinfo {volume} {68}},\ \bibinfo {pages} {021304} (\bibinfo {year} {2003})},\ \Eprint {http://arxiv.org/abs/hep-ph/0304261} {arXiv:hep-ph/0304261} \BibitemShut {NoStop}%
\bibitem [{\citenamefont {An}\ \emph {et~al.}(2010)\citenamefont {An}, \citenamefont {Chen}, \citenamefont {Mohapatra},\ and\ \citenamefont {Zhang}}]{An:2009vq}%
  \BibitemOpen
  \bibfield  {author} {\bibinfo {author} {\bibfnamefont {H.}~\bibnamefont {An}}, \bibinfo {author} {\bibfnamefont {S.-L.}\ \bibnamefont {Chen}}, \bibinfo {author} {\bibfnamefont {R.~N.}\ \bibnamefont {Mohapatra}}, \ and\ \bibinfo {author} {\bibfnamefont {Y.}~\bibnamefont {Zhang}},\ }\href {\doibase 10.1007/JHEP03(2010)124} {\bibfield  {journal} {\bibinfo  {journal} {JHEP}\ }\textbf {\bibinfo {volume} {03}},\ \bibinfo {pages} {124} (\bibinfo {year} {2010})},\ \Eprint {http://arxiv.org/abs/0911.4463} {arXiv:0911.4463 [hep-ph]} \BibitemShut {NoStop}%
\bibitem [{\citenamefont {Farina}(2015)}]{Farina:2015uea}%
  \BibitemOpen
  \bibfield  {author} {\bibinfo {author} {\bibfnamefont {M.}~\bibnamefont {Farina}},\ }\href {\doibase 10.1088/1475-7516/2015/11/017} {\bibfield  {journal} {\bibinfo  {journal} {JCAP}\ }\textbf {\bibinfo {volume} {11}},\ \bibinfo {pages} {017} (\bibinfo {year} {2015})},\ \Eprint {http://arxiv.org/abs/1506.03520} {arXiv:1506.03520 [hep-ph]} \BibitemShut {NoStop}%
\bibitem [{\citenamefont {Garcia~Garcia}\ \emph {et~al.}(2015)\citenamefont {Garcia~Garcia}, \citenamefont {Lasenby},\ and\ \citenamefont {March-Russell}}]{GarciaGarcia:2015pnn}%
  \BibitemOpen
  \bibfield  {author} {\bibinfo {author} {\bibfnamefont {I.}~\bibnamefont {Garcia~Garcia}}, \bibinfo {author} {\bibfnamefont {R.}~\bibnamefont {Lasenby}}, \ and\ \bibinfo {author} {\bibfnamefont {J.}~\bibnamefont {March-Russell}},\ }\href {\doibase 10.1103/PhysRevLett.115.121801} {\bibfield  {journal} {\bibinfo  {journal} {Phys. Rev. Lett.}\ }\textbf {\bibinfo {volume} {115}},\ \bibinfo {pages} {121801} (\bibinfo {year} {2015})},\ \Eprint {http://arxiv.org/abs/1505.07410} {arXiv:1505.07410 [hep-ph]} \BibitemShut {NoStop}%
\bibitem [{\citenamefont {Lonsdale}\ and\ \citenamefont {Volkas}(2018)}]{Lonsdale:2018xwd}%
  \BibitemOpen
  \bibfield  {author} {\bibinfo {author} {\bibfnamefont {S.~J.}\ \bibnamefont {Lonsdale}}\ and\ \bibinfo {author} {\bibfnamefont {R.~R.}\ \bibnamefont {Volkas}},\ }\href {\doibase 10.1103/PhysRevD.97.103510} {\bibfield  {journal} {\bibinfo  {journal} {Phys. Rev. D}\ }\textbf {\bibinfo {volume} {97}},\ \bibinfo {pages} {103510} (\bibinfo {year} {2018})},\ \Eprint {http://arxiv.org/abs/1801.05561} {arXiv:1801.05561 [hep-ph]} \BibitemShut {NoStop}%
\bibitem [{\citenamefont {Ibe}\ \emph {et~al.}(2019)\citenamefont {Ibe}, \citenamefont {Kamada}, \citenamefont {Kobayashi}, \citenamefont {Kuwahara},\ and\ \citenamefont {Nakano}}]{Ibe:2019ena}%
  \BibitemOpen
  \bibfield  {author} {\bibinfo {author} {\bibfnamefont {M.}~\bibnamefont {Ibe}}, \bibinfo {author} {\bibfnamefont {A.}~\bibnamefont {Kamada}}, \bibinfo {author} {\bibfnamefont {S.}~\bibnamefont {Kobayashi}}, \bibinfo {author} {\bibfnamefont {T.}~\bibnamefont {Kuwahara}}, \ and\ \bibinfo {author} {\bibfnamefont {W.}~\bibnamefont {Nakano}},\ }\href {\doibase 10.1103/PhysRevD.100.075022} {\bibfield  {journal} {\bibinfo  {journal} {Phys. Rev. D}\ }\textbf {\bibinfo {volume} {100}},\ \bibinfo {pages} {075022} (\bibinfo {year} {2019})},\ \Eprint {http://arxiv.org/abs/1907.03404} {arXiv:1907.03404 [hep-ph]} \BibitemShut {NoStop}%
\bibitem [{\citenamefont {Bodas}\ \emph {et~al.}(2024)\citenamefont {Bodas}, \citenamefont {Buen-Abad}, \citenamefont {Hook},\ and\ \citenamefont {Sundrum}}]{Bodas:2024idn}%
  \BibitemOpen
  \bibfield  {author} {\bibinfo {author} {\bibfnamefont {A.}~\bibnamefont {Bodas}}, \bibinfo {author} {\bibfnamefont {M.~A.}\ \bibnamefont {Buen-Abad}}, \bibinfo {author} {\bibfnamefont {A.}~\bibnamefont {Hook}}, \ and\ \bibinfo {author} {\bibfnamefont {R.}~\bibnamefont {Sundrum}},\ }\href@noop {} {\  (\bibinfo {year} {2024})},\ \Eprint {http://arxiv.org/abs/2401.12286} {arXiv:2401.12286 [hep-ph]} \BibitemShut {NoStop}%
\bibitem [{\citenamefont {Bai}\ and\ \citenamefont {Schwaller}(2014)}]{Bai:2013xga}%
  \BibitemOpen
  \bibfield  {author} {\bibinfo {author} {\bibfnamefont {Y.}~\bibnamefont {Bai}}\ and\ \bibinfo {author} {\bibfnamefont {P.}~\bibnamefont {Schwaller}},\ }\href {\doibase 10.1103/PhysRevD.89.063522} {\bibfield  {journal} {\bibinfo  {journal} {Phys. Rev. D}\ }\textbf {\bibinfo {volume} {89}},\ \bibinfo {pages} {063522} (\bibinfo {year} {2014})},\ \Eprint {http://arxiv.org/abs/1306.4676} {arXiv:1306.4676 [hep-ph]} \BibitemShut {NoStop}%
\bibitem [{\citenamefont {Newstead}\ and\ \citenamefont {TerBeek}(2014)}]{Newstead:2014jva}%
  \BibitemOpen
  \bibfield  {author} {\bibinfo {author} {\bibfnamefont {J.~L.}\ \bibnamefont {Newstead}}\ and\ \bibinfo {author} {\bibfnamefont {R.~H.}\ \bibnamefont {TerBeek}},\ }\href {\doibase 10.1103/PhysRevD.90.074008} {\bibfield  {journal} {\bibinfo  {journal} {Phys. Rev. D}\ }\textbf {\bibinfo {volume} {90}},\ \bibinfo {pages} {074008} (\bibinfo {year} {2014})},\ \Eprint {http://arxiv.org/abs/1405.7427} {arXiv:1405.7427 [hep-ph]} \BibitemShut {NoStop}%
\bibitem [{\citenamefont {Ritter}\ and\ \citenamefont {Volkas}(2023)}]{Ritter:2022opo}%
  \BibitemOpen
  \bibfield  {author} {\bibinfo {author} {\bibfnamefont {A.~C.}\ \bibnamefont {Ritter}}\ and\ \bibinfo {author} {\bibfnamefont {R.~R.}\ \bibnamefont {Volkas}},\ }\href {\doibase 10.1103/PhysRevD.107.015029} {\bibfield  {journal} {\bibinfo  {journal} {Phys. Rev. D}\ }\textbf {\bibinfo {volume} {107}},\ \bibinfo {pages} {015029} (\bibinfo {year} {2023})},\ \Eprint {http://arxiv.org/abs/2210.11011} {arXiv:2210.11011 [hep-ph]} \BibitemShut {NoStop}%
\bibitem [{\citenamefont {Ritter}\ and\ \citenamefont {Volkas}(2024)}]{Ritter:2024sqv}%
  \BibitemOpen
  \bibfield  {author} {\bibinfo {author} {\bibfnamefont {A.~C.}\ \bibnamefont {Ritter}}\ and\ \bibinfo {author} {\bibfnamefont {R.~R.}\ \bibnamefont {Volkas}},\ }\href@noop {} {\  (\bibinfo {year} {2024})},\ \Eprint {http://arxiv.org/abs/2404.05999} {arXiv:2404.05999 [hep-ph]} \BibitemShut {NoStop}%
\bibitem [{\citenamefont {Murgui}\ and\ \citenamefont {Zurek}(2022)}]{Murgui:2021eqf}%
  \BibitemOpen
  \bibfield  {author} {\bibinfo {author} {\bibfnamefont {C.}~\bibnamefont {Murgui}}\ and\ \bibinfo {author} {\bibfnamefont {K.~M.}\ \bibnamefont {Zurek}},\ }\href {\doibase 10.1103/PhysRevD.105.095002} {\bibfield  {journal} {\bibinfo  {journal} {Phys. Rev. D}\ }\textbf {\bibinfo {volume} {105}},\ \bibinfo {pages} {095002} (\bibinfo {year} {2022})},\ \Eprint {http://arxiv.org/abs/2112.08374} {arXiv:2112.08374 [hep-ph]} \BibitemShut {NoStop}%
\bibitem [{\citenamefont {Chung}(2024{\natexlab{a}})}]{Chung:2024nnj}%
  \BibitemOpen
  \bibfield  {author} {\bibinfo {author} {\bibfnamefont {Y.}~\bibnamefont {Chung}},\ }\href@noop {} {\  (\bibinfo {year} {2024}{\natexlab{a}})},\ \Eprint {http://arxiv.org/abs/2411.16860} {arXiv:2411.16860 [hep-ph]} \BibitemShut {NoStop}%
\bibitem [{\citenamefont {Rosa}\ and\ \citenamefont {Silva}(2023)}]{Rosa:2022sym}%
  \BibitemOpen
  \bibfield  {author} {\bibinfo {author} {\bibfnamefont {J.~a.~G.}\ \bibnamefont {Rosa}}\ and\ \bibinfo {author} {\bibfnamefont {D.~M.~C.}\ \bibnamefont {Silva}},\ }\href {\doibase 10.1016/j.physletb.2023.138178} {\bibfield  {journal} {\bibinfo  {journal} {Phys. Lett. B}\ }\textbf {\bibinfo {volume} {846}},\ \bibinfo {pages} {138178} (\bibinfo {year} {2023})},\ \Eprint {http://arxiv.org/abs/2211.10359} {arXiv:2211.10359 [hep-ph]} \BibitemShut {NoStop}%
\bibitem [{\citenamefont {Brzeminski}\ and\ \citenamefont {Hook}(2024)}]{Brzeminski:2023wza}%
  \BibitemOpen
  \bibfield  {author} {\bibinfo {author} {\bibfnamefont {D.}~\bibnamefont {Brzeminski}}\ and\ \bibinfo {author} {\bibfnamefont {A.}~\bibnamefont {Hook}},\ }\href {\doibase 10.1103/PhysRevLett.132.201001} {\bibfield  {journal} {\bibinfo  {journal} {Phys. Rev. Lett.}\ }\textbf {\bibinfo {volume} {132}},\ \bibinfo {pages} {201001} (\bibinfo {year} {2024})},\ \Eprint {http://arxiv.org/abs/2310.07777} {arXiv:2310.07777 [hep-ph]} \BibitemShut {NoStop}%
\bibitem [{\citenamefont {Banerjee}\ \emph {et~al.}(2024)\citenamefont {Banerjee}, \citenamefont {Brzeminski},\ and\ \citenamefont {Hook}}]{Banerjee:2024xhn}%
  \BibitemOpen
  \bibfield  {author} {\bibinfo {author} {\bibfnamefont {A.}~\bibnamefont {Banerjee}}, \bibinfo {author} {\bibfnamefont {D.}~\bibnamefont {Brzeminski}}, \ and\ \bibinfo {author} {\bibfnamefont {A.}~\bibnamefont {Hook}},\ }\href@noop {} {\  (\bibinfo {year} {2024})},\ \Eprint {http://arxiv.org/abs/2410.22412} {arXiv:2410.22412 [hep-ph]} \BibitemShut {NoStop}%
\bibitem [{\citenamefont {Thomas}\ and\ \citenamefont {Xu}(1992)}]{Thomas:1992hf}%
  \BibitemOpen
  \bibfield  {author} {\bibinfo {author} {\bibfnamefont {S.~D.}\ \bibnamefont {Thomas}}\ and\ \bibinfo {author} {\bibfnamefont {R.-M.}\ \bibnamefont {Xu}},\ }\href {\doibase 10.1016/0370-2693(92)90442-7} {\bibfield  {journal} {\bibinfo  {journal} {Phys. Lett. B}\ }\textbf {\bibinfo {volume} {284}},\ \bibinfo {pages} {341} (\bibinfo {year} {1992})}\BibitemShut {NoStop}%
\bibitem [{\citenamefont {McDonald}(1996)}]{McDonald:1996cs}%
  \BibitemOpen
  \bibfield  {author} {\bibinfo {author} {\bibfnamefont {J.}~\bibnamefont {McDonald}},\ }\href@noop {} {\  (\bibinfo {year} {1996})},\ \Eprint {http://arxiv.org/abs/hep-ph/9610324} {arXiv:hep-ph/9610324} \BibitemShut {NoStop}%
\bibitem [{\citenamefont {Ibanez}\ \emph {et~al.}(2001)\citenamefont {Ibanez}, \citenamefont {Marchesano},\ and\ \citenamefont {Rabadan}}]{Ibanez:2001nd}%
  \BibitemOpen
  \bibfield  {author} {\bibinfo {author} {\bibfnamefont {L.~E.}\ \bibnamefont {Ibanez}}, \bibinfo {author} {\bibfnamefont {F.}~\bibnamefont {Marchesano}}, \ and\ \bibinfo {author} {\bibfnamefont {R.}~\bibnamefont {Rabadan}},\ }\href {\doibase 10.1088/1126-6708/2001/11/002} {\bibfield  {journal} {\bibinfo  {journal} {JHEP}\ }\textbf {\bibinfo {volume} {11}},\ \bibinfo {pages} {002} (\bibinfo {year} {2001})},\ \Eprint {http://arxiv.org/abs/hep-th/0105155} {arXiv:hep-th/0105155} \BibitemShut {NoStop}%
\bibitem [{\citenamefont {Davoudiasl}\ and\ \citenamefont {Everett}(2006)}]{Davoudiasl:2005ai}%
  \BibitemOpen
  \bibfield  {author} {\bibinfo {author} {\bibfnamefont {H.}~\bibnamefont {Davoudiasl}}\ and\ \bibinfo {author} {\bibfnamefont {L.~L.}\ \bibnamefont {Everett}},\ }\href {\doibase 10.1016/j.physletb.2006.01.039} {\bibfield  {journal} {\bibinfo  {journal} {Phys. Lett. B}\ }\textbf {\bibinfo {volume} {634}},\ \bibinfo {pages} {55} (\bibinfo {year} {2006})},\ \Eprint {http://arxiv.org/abs/hep-ph/0512188} {arXiv:hep-ph/0512188} \BibitemShut {NoStop}%
\bibitem [{\citenamefont {Babi\v{c}}\ \emph {et~al.}(2021)\citenamefont {Babi\v{c}}, \citenamefont {Kovalenko}, \citenamefont {Krivoruchenko},\ and\ \citenamefont {\v{S}imkovic}}]{Babic:2019zqu}%
  \BibitemOpen
  \bibfield  {author} {\bibinfo {author} {\bibfnamefont {A.}~\bibnamefont {Babi\v{c}}}, \bibinfo {author} {\bibfnamefont {S.}~\bibnamefont {Kovalenko}}, \bibinfo {author} {\bibfnamefont {M.~I.}\ \bibnamefont {Krivoruchenko}}, \ and\ \bibinfo {author} {\bibfnamefont {F.}~\bibnamefont {\v{S}imkovic}},\ }\href {\doibase 10.1103/PhysRevD.103.015007} {\bibfield  {journal} {\bibinfo  {journal} {Phys. Rev. D}\ }\textbf {\bibinfo {volume} {103}},\ \bibinfo {pages} {015007} (\bibinfo {year} {2021})},\ \Eprint {http://arxiv.org/abs/1911.12189} {arXiv:1911.12189 [hep-ph]} \BibitemShut {NoStop}%
\bibitem [{\citenamefont {Davoudiasl}\ \emph {et~al.}(2022)\citenamefont {Davoudiasl}, \citenamefont {Lewis},\ and\ \citenamefont {Sullivan}}]{Davoudiasl:2021nfv}%
  \BibitemOpen
  \bibfield  {author} {\bibinfo {author} {\bibfnamefont {H.}~\bibnamefont {Davoudiasl}}, \bibinfo {author} {\bibfnamefont {I.~M.}\ \bibnamefont {Lewis}}, \ and\ \bibinfo {author} {\bibfnamefont {M.}~\bibnamefont {Sullivan}},\ }\href {\doibase 10.1103/PhysRevD.105.075017} {\bibfield  {journal} {\bibinfo  {journal} {Phys. Rev. D}\ }\textbf {\bibinfo {volume} {105}},\ \bibinfo {pages} {075017} (\bibinfo {year} {2022})},\ \Eprint {http://arxiv.org/abs/2111.08020} {arXiv:2111.08020 [hep-ph]} \BibitemShut {NoStop}%
\bibitem [{\citenamefont {Pati}\ and\ \citenamefont {Salam}(1974)}]{Pati:1974yy}%
  \BibitemOpen
  \bibfield  {author} {\bibinfo {author} {\bibfnamefont {J.~C.}\ \bibnamefont {Pati}}\ and\ \bibinfo {author} {\bibfnamefont {A.}~\bibnamefont {Salam}},\ }\href {\doibase 10.1103/PhysRevD.10.275} {\bibfield  {journal} {\bibinfo  {journal} {Phys. Rev. D}\ }\textbf {\bibinfo {volume} {10}},\ \bibinfo {pages} {275} (\bibinfo {year} {1974})},\ \bibinfo {note} {[Erratum: Phys.Rev.D 11, 703--703 (1975)]}\BibitemShut {NoStop}%
\bibitem [{\citenamefont {Weinberg}(1979)}]{Weinberg:1978kz}%
  \BibitemOpen
  \bibfield  {author} {\bibinfo {author} {\bibfnamefont {S.}~\bibnamefont {Weinberg}},\ }\href {\doibase 10.1016/0378-4371(79)90223-1} {\bibfield  {journal} {\bibinfo  {journal} {Physica A}\ }\textbf {\bibinfo {volume} {96}},\ \bibinfo {pages} {327} (\bibinfo {year} {1979})}\BibitemShut {NoStop}%
\bibitem [{\citenamefont {Manohar}\ and\ \citenamefont {Georgi}(1984)}]{Manohar:1983md}%
  \BibitemOpen
  \bibfield  {author} {\bibinfo {author} {\bibfnamefont {A.}~\bibnamefont {Manohar}}\ and\ \bibinfo {author} {\bibfnamefont {H.}~\bibnamefont {Georgi}},\ }\href {\doibase 10.1016/0550-3213(84)90231-1} {\bibfield  {journal} {\bibinfo  {journal} {Nucl. Phys. B}\ }\textbf {\bibinfo {volume} {234}},\ \bibinfo {pages} {189} (\bibinfo {year} {1984})}\BibitemShut {NoStop}%
\bibitem [{\citenamefont {Georgi}(1986)}]{Georgi:1985hf}%
  \BibitemOpen
  \bibfield  {author} {\bibinfo {author} {\bibfnamefont {H.}~\bibnamefont {Georgi}},\ }\href {\doibase 10.1016/0550-3213(86)90092-1} {\bibfield  {journal} {\bibinfo  {journal} {Nucl. Phys. B}\ }\textbf {\bibinfo {volume} {266}},\ \bibinfo {pages} {274} (\bibinfo {year} {1986})}\BibitemShut {NoStop}%
\bibitem [{\citenamefont {Kaplan}\ and\ \citenamefont {Georgi}(1984)}]{Kaplan:1983fs}%
  \BibitemOpen
  \bibfield  {author} {\bibinfo {author} {\bibfnamefont {D.~B.}\ \bibnamefont {Kaplan}}\ and\ \bibinfo {author} {\bibfnamefont {H.}~\bibnamefont {Georgi}},\ }\href {\doibase 10.1016/0370-2693(84)91177-8} {\bibfield  {journal} {\bibinfo  {journal} {Phys. Lett.}\ }\textbf {\bibinfo {volume} {136B}},\ \bibinfo {pages} {183} (\bibinfo {year} {1984})}\BibitemShut {NoStop}%
\bibitem [{\citenamefont {Kaplan}\ \emph {et~al.}(1984)\citenamefont {Kaplan}, \citenamefont {Georgi},\ and\ \citenamefont {Dimopoulos}}]{Kaplan:1983sm}%
  \BibitemOpen
  \bibfield  {author} {\bibinfo {author} {\bibfnamefont {D.~B.}\ \bibnamefont {Kaplan}}, \bibinfo {author} {\bibfnamefont {H.}~\bibnamefont {Georgi}}, \ and\ \bibinfo {author} {\bibfnamefont {S.}~\bibnamefont {Dimopoulos}},\ }\href {\doibase 10.1016/0370-2693(84)91178-X} {\bibfield  {journal} {\bibinfo  {journal} {Phys. Lett.}\ }\textbf {\bibinfo {volume} {136B}},\ \bibinfo {pages} {187} (\bibinfo {year} {1984})}\BibitemShut {NoStop}%
\bibitem [{\citenamefont {Bellazzini}\ \emph {et~al.}(2014)\citenamefont {Bellazzini}, \citenamefont {Csáki},\ and\ \citenamefont {Serra}}]{Bellazzini:2014yua}%
  \BibitemOpen
  \bibfield  {author} {\bibinfo {author} {\bibfnamefont {B.}~\bibnamefont {Bellazzini}}, \bibinfo {author} {\bibfnamefont {C.}~\bibnamefont {Csáki}}, \ and\ \bibinfo {author} {\bibfnamefont {J.}~\bibnamefont {Serra}},\ }\href {\doibase 10.1140/epjc/s10052-014-2766-x} {\bibfield  {journal} {\bibinfo  {journal} {Eur. Phys. J.}\ }\textbf {\bibinfo {volume} {C74}},\ \bibinfo {pages} {2766} (\bibinfo {year} {2014})},\ \Eprint {http://arxiv.org/abs/1401.2457} {arXiv:1401.2457 [hep-ph]} \BibitemShut {NoStop}%
\bibitem [{\citenamefont {Panico}\ and\ \citenamefont {Wulzer}(2016)}]{Panico:2015jxa}%
  \BibitemOpen
  \bibfield  {author} {\bibinfo {author} {\bibfnamefont {G.}~\bibnamefont {Panico}}\ and\ \bibinfo {author} {\bibfnamefont {A.}~\bibnamefont {Wulzer}},\ }\href {\doibase 10.1007/978-3-319-22617-0} {\bibfield  {journal} {\bibinfo  {journal} {Lect. Notes Phys.}\ }\textbf {\bibinfo {volume} {913}},\ \bibinfo {pages} {pp.1} (\bibinfo {year} {2016})},\ \Eprint {http://arxiv.org/abs/1506.01961} {arXiv:1506.01961 [hep-ph]} \BibitemShut {NoStop}%
\bibitem [{\citenamefont {Cacciapaglia}\ and\ \citenamefont {Sannino}(2014)}]{Cacciapaglia:2014uja}%
  \BibitemOpen
  \bibfield  {author} {\bibinfo {author} {\bibfnamefont {G.}~\bibnamefont {Cacciapaglia}}\ and\ \bibinfo {author} {\bibfnamefont {F.}~\bibnamefont {Sannino}},\ }\href {\doibase 10.1007/JHEP04(2014)111} {\bibfield  {journal} {\bibinfo  {journal} {JHEP}\ }\textbf {\bibinfo {volume} {04}},\ \bibinfo {pages} {111} (\bibinfo {year} {2014})},\ \Eprint {http://arxiv.org/abs/1402.0233} {arXiv:1402.0233 [hep-ph]} \BibitemShut {NoStop}%
\bibitem [{\citenamefont {Cacciapaglia}\ \emph {et~al.}(2020)\citenamefont {Cacciapaglia}, \citenamefont {Pica},\ and\ \citenamefont {Sannino}}]{Cacciapaglia:2020kgq}%
  \BibitemOpen
  \bibfield  {author} {\bibinfo {author} {\bibfnamefont {G.}~\bibnamefont {Cacciapaglia}}, \bibinfo {author} {\bibfnamefont {C.}~\bibnamefont {Pica}}, \ and\ \bibinfo {author} {\bibfnamefont {F.}~\bibnamefont {Sannino}},\ }\href {\doibase 10.1016/j.physrep.2020.07.002} {\bibfield  {journal} {\bibinfo  {journal} {Phys. Rept.}\ }\textbf {\bibinfo {volume} {877}},\ \bibinfo {pages} {1} (\bibinfo {year} {2020})},\ \Eprint {http://arxiv.org/abs/2002.04914} {arXiv:2002.04914 [hep-ph]} \BibitemShut {NoStop}%
\bibitem [{\citenamefont {Cacciapaglia}\ and\ \citenamefont {Sannino}(2016)}]{Cacciapaglia:2015yra}%
  \BibitemOpen
  \bibfield  {author} {\bibinfo {author} {\bibfnamefont {G.}~\bibnamefont {Cacciapaglia}}\ and\ \bibinfo {author} {\bibfnamefont {F.}~\bibnamefont {Sannino}},\ }\href {\doibase 10.1016/j.physletb.2016.02.034} {\bibfield  {journal} {\bibinfo  {journal} {Phys. Lett. B}\ }\textbf {\bibinfo {volume} {755}},\ \bibinfo {pages} {328} (\bibinfo {year} {2016})},\ \Eprint {http://arxiv.org/abs/1508.00016} {arXiv:1508.00016 [hep-ph]} \BibitemShut {NoStop}%
\bibitem [{\citenamefont {Chung}(2024{\natexlab{b}})}]{Chung:2023iwj}%
  \BibitemOpen
  \bibfield  {author} {\bibinfo {author} {\bibfnamefont {Y.}~\bibnamefont {Chung}},\ }\href {\doibase 10.1103/PhysRevD.109.095021} {\bibfield  {journal} {\bibinfo  {journal} {Phys. Rev. D}\ }\textbf {\bibinfo {volume} {109}},\ \bibinfo {pages} {095021} (\bibinfo {year} {2024}{\natexlab{b}})},\ \Eprint {http://arxiv.org/abs/2309.00072} {arXiv:2309.00072 [hep-ph]} \BibitemShut {NoStop}%
\bibitem [{\citenamefont {Dimopoulos}\ \emph {et~al.}(1980)\citenamefont {Dimopoulos}, \citenamefont {Raby},\ and\ \citenamefont {Susskind}}]{Dimopoulos:1980hn}%
  \BibitemOpen
  \bibfield  {author} {\bibinfo {author} {\bibfnamefont {S.}~\bibnamefont {Dimopoulos}}, \bibinfo {author} {\bibfnamefont {S.}~\bibnamefont {Raby}}, \ and\ \bibinfo {author} {\bibfnamefont {L.}~\bibnamefont {Susskind}},\ }\href {\doibase 10.1016/0550-3213(80)90215-1} {\bibfield  {journal} {\bibinfo  {journal} {Nucl. Phys. B}\ }\textbf {\bibinfo {volume} {173}},\ \bibinfo {pages} {208} (\bibinfo {year} {1980})}\BibitemShut {NoStop}%
\bibitem [{\citenamefont {Raby}\ \emph {et~al.}(1980)\citenamefont {Raby}, \citenamefont {Dimopoulos},\ and\ \citenamefont {Susskind}}]{Raby:1979my}%
  \BibitemOpen
  \bibfield  {author} {\bibinfo {author} {\bibfnamefont {S.}~\bibnamefont {Raby}}, \bibinfo {author} {\bibfnamefont {S.}~\bibnamefont {Dimopoulos}}, \ and\ \bibinfo {author} {\bibfnamefont {L.}~\bibnamefont {Susskind}},\ }\href {\doibase 10.1016/0550-3213(80)90093-0} {\bibfield  {journal} {\bibinfo  {journal} {Nucl. Phys. B}\ }\textbf {\bibinfo {volume} {169}},\ \bibinfo {pages} {373} (\bibinfo {year} {1980})}\BibitemShut {NoStop}%
\bibitem [{\citenamefont {Bars}\ and\ \citenamefont {Yankielowicz}(1981)}]{Bars:1981se}%
  \BibitemOpen
  \bibfield  {author} {\bibinfo {author} {\bibfnamefont {I.}~\bibnamefont {Bars}}\ and\ \bibinfo {author} {\bibfnamefont {S.}~\bibnamefont {Yankielowicz}},\ }\href {\doibase 10.1016/0370-2693(81)90664-X} {\bibfield  {journal} {\bibinfo  {journal} {Phys. Lett. B}\ }\textbf {\bibinfo {volume} {101}},\ \bibinfo {pages} {159} (\bibinfo {year} {1981})}\BibitemShut {NoStop}%
\bibitem [{\citenamefont {Appelquist}\ \emph {et~al.}(2000)\citenamefont {Appelquist}, \citenamefont {Duan},\ and\ \citenamefont {Sannino}}]{Appelquist:2000qg}%
  \BibitemOpen
  \bibfield  {author} {\bibinfo {author} {\bibfnamefont {T.}~\bibnamefont {Appelquist}}, \bibinfo {author} {\bibfnamefont {Z.-y.}\ \bibnamefont {Duan}}, \ and\ \bibinfo {author} {\bibfnamefont {F.}~\bibnamefont {Sannino}},\ }\href {\doibase 10.1103/PhysRevD.61.125009} {\bibfield  {journal} {\bibinfo  {journal} {Phys. Rev. D}\ }\textbf {\bibinfo {volume} {61}},\ \bibinfo {pages} {125009} (\bibinfo {year} {2000})},\ \Eprint {http://arxiv.org/abs/hep-ph/0001043} {arXiv:hep-ph/0001043} \BibitemShut {NoStop}%
\bibitem [{\citenamefont {Cs\'aki}\ \emph {et~al.}(2021)\citenamefont {Cs\'aki}, \citenamefont {Murayama},\ and\ \citenamefont {Telem}}]{Csaki:2021xhi}%
  \BibitemOpen
  \bibfield  {author} {\bibinfo {author} {\bibfnamefont {C.}~\bibnamefont {Cs\'aki}}, \bibinfo {author} {\bibfnamefont {H.}~\bibnamefont {Murayama}}, \ and\ \bibinfo {author} {\bibfnamefont {O.}~\bibnamefont {Telem}},\ }\href {\doibase 10.1103/PhysRevD.104.065018} {\bibfield  {journal} {\bibinfo  {journal} {Phys. Rev. D}\ }\textbf {\bibinfo {volume} {104}},\ \bibinfo {pages} {065018} (\bibinfo {year} {2021})},\ \Eprint {http://arxiv.org/abs/2104.10171} {arXiv:2104.10171 [hep-th]} \BibitemShut {NoStop}%
\bibitem [{\citenamefont {Bolognesi}\ \emph {et~al.}(2022)\citenamefont {Bolognesi}, \citenamefont {Konishi},\ and\ \citenamefont {Luzio}}]{Bolognesi:2021jzs}%
  \BibitemOpen
  \bibfield  {author} {\bibinfo {author} {\bibfnamefont {S.}~\bibnamefont {Bolognesi}}, \bibinfo {author} {\bibfnamefont {K.}~\bibnamefont {Konishi}}, \ and\ \bibinfo {author} {\bibfnamefont {A.}~\bibnamefont {Luzio}},\ }\href {\doibase 10.1142/S0217751X22300149} {\bibfield  {journal} {\bibinfo  {journal} {Int. J. Mod. Phys. A}\ }\textbf {\bibinfo {volume} {37}},\ \bibinfo {pages} {2230014} (\bibinfo {year} {2022})},\ \Eprint {http://arxiv.org/abs/2110.02104} {arXiv:2110.02104 [hep-th]} \BibitemShut {NoStop}%
\bibitem [{\citenamefont {Smith}\ \emph {et~al.}(2022)\citenamefont {Smith}, \citenamefont {Karasik}, \citenamefont {Lohitsiri},\ and\ \citenamefont {Tong}}]{Smith:2021vbf}%
  \BibitemOpen
  \bibfield  {author} {\bibinfo {author} {\bibfnamefont {P.~B.}\ \bibnamefont {Smith}}, \bibinfo {author} {\bibfnamefont {A.}~\bibnamefont {Karasik}}, \bibinfo {author} {\bibfnamefont {N.}~\bibnamefont {Lohitsiri}}, \ and\ \bibinfo {author} {\bibfnamefont {D.}~\bibnamefont {Tong}},\ }\href {\doibase 10.1007/JHEP01(2022)112} {\bibfield  {journal} {\bibinfo  {journal} {JHEP}\ }\textbf {\bibinfo {volume} {01}},\ \bibinfo {pages} {112} (\bibinfo {year} {2022})},\ \Eprint {http://arxiv.org/abs/2106.06402} {arXiv:2106.06402 [hep-th]} \BibitemShut {NoStop}%
\bibitem [{\citenamefont {Karasik}\ \emph {et~al.}(2022)\citenamefont {Karasik}, \citenamefont {\"Onder},\ and\ \citenamefont {Tong}}]{Karasik:2022gve}%
  \BibitemOpen
  \bibfield  {author} {\bibinfo {author} {\bibfnamefont {A.}~\bibnamefont {Karasik}}, \bibinfo {author} {\bibfnamefont {K.}~\bibnamefont {\"Onder}}, \ and\ \bibinfo {author} {\bibfnamefont {D.}~\bibnamefont {Tong}},\ }\href {\doibase 10.1007/JHEP11(2022)122} {\bibfield  {journal} {\bibinfo  {journal} {JHEP}\ }\textbf {\bibinfo {volume} {11}},\ \bibinfo {pages} {122} (\bibinfo {year} {2022})},\ \Eprint {http://arxiv.org/abs/2208.07842} {arXiv:2208.07842 [hep-th]} \BibitemShut {NoStop}%
\bibitem [{\citenamefont {Li}\ \emph {et~al.}(2025)\citenamefont {Li}, \citenamefont {Pastor-Guti{\'e}rrez}, \citenamefont {Vatani},\ and\ \citenamefont {Xu}}]{Li:2025tvu}%
  \BibitemOpen
  \bibfield  {author} {\bibinfo {author} {\bibfnamefont {H.-L.}\ \bibnamefont {Li}}, \bibinfo {author} {\bibfnamefont {{\'A}.}~\bibnamefont {Pastor-Guti{\'e}rrez}}, \bibinfo {author} {\bibfnamefont {S.}~\bibnamefont {Vatani}}, \ and\ \bibinfo {author} {\bibfnamefont {L.-X.}\ \bibnamefont {Xu}},\ }\href@noop {} {\  (\bibinfo {year} {2025})},\ \Eprint {http://arxiv.org/abs/2507.21208} {arXiv:2507.21208 [hep-th]} \BibitemShut {NoStop}%
\bibitem [{\citenamefont {Kim}(1985)}]{Kim:1984pt}%
  \BibitemOpen
  \bibfield  {author} {\bibinfo {author} {\bibfnamefont {J.~E.}\ \bibnamefont {Kim}},\ }\href {\doibase 10.1103/PhysRevD.31.1733} {\bibfield  {journal} {\bibinfo  {journal} {Phys. Rev. D}\ }\textbf {\bibinfo {volume} {31}},\ \bibinfo {pages} {1733} (\bibinfo {year} {1985})}\BibitemShut {NoStop}%
\bibitem [{\citenamefont {Choi}\ and\ \citenamefont {Kim}(1985)}]{Choi:1985cb}%
  \BibitemOpen
  \bibfield  {author} {\bibinfo {author} {\bibfnamefont {K.}~\bibnamefont {Choi}}\ and\ \bibinfo {author} {\bibfnamefont {J.~E.}\ \bibnamefont {Kim}},\ }\href {\doibase 10.1103/PhysRevD.32.1828} {\bibfield  {journal} {\bibinfo  {journal} {Phys. Rev. D}\ }\textbf {\bibinfo {volume} {32}},\ \bibinfo {pages} {1828} (\bibinfo {year} {1985})}\BibitemShut {NoStop}%
\bibitem [{\citenamefont {Kaplan}(1985)}]{Kaplan:1985dv}%
  \BibitemOpen
  \bibfield  {author} {\bibinfo {author} {\bibfnamefont {D.~B.}\ \bibnamefont {Kaplan}},\ }\href {\doibase 10.1016/0550-3213(85)90319-0} {\bibfield  {journal} {\bibinfo  {journal} {Nucl. Phys. B}\ }\textbf {\bibinfo {volume} {260}},\ \bibinfo {pages} {215} (\bibinfo {year} {1985})}\BibitemShut {NoStop}%
\bibitem [{\citenamefont {Aloni}\ \emph {et~al.}(2019)\citenamefont {Aloni}, \citenamefont {Soreq},\ and\ \citenamefont {Williams}}]{Aloni:2018vki}%
  \BibitemOpen
  \bibfield  {author} {\bibinfo {author} {\bibfnamefont {D.}~\bibnamefont {Aloni}}, \bibinfo {author} {\bibfnamefont {Y.}~\bibnamefont {Soreq}}, \ and\ \bibinfo {author} {\bibfnamefont {M.}~\bibnamefont {Williams}},\ }\href {\doibase 10.1103/PhysRevLett.123.031803} {\bibfield  {journal} {\bibinfo  {journal} {Phys. Rev. Lett.}\ }\textbf {\bibinfo {volume} {123}},\ \bibinfo {pages} {031803} (\bibinfo {year} {2019})},\ \Eprint {http://arxiv.org/abs/1811.03474} {arXiv:1811.03474 [hep-ph]} \BibitemShut {NoStop}%
\bibitem [{\citenamefont {Bauer}\ \emph {et~al.}(2017)\citenamefont {Bauer}, \citenamefont {Neubert},\ and\ \citenamefont {Thamm}}]{Bauer:2017ris}%
  \BibitemOpen
  \bibfield  {author} {\bibinfo {author} {\bibfnamefont {M.}~\bibnamefont {Bauer}}, \bibinfo {author} {\bibfnamefont {M.}~\bibnamefont {Neubert}}, \ and\ \bibinfo {author} {\bibfnamefont {A.}~\bibnamefont {Thamm}},\ }\href {\doibase 10.1007/JHEP12(2017)044} {\bibfield  {journal} {\bibinfo  {journal} {JHEP}\ }\textbf {\bibinfo {volume} {12}},\ \bibinfo {pages} {044} (\bibinfo {year} {2017})},\ \Eprint {http://arxiv.org/abs/1708.00443} {arXiv:1708.00443 [hep-ph]} \BibitemShut {NoStop}%
\bibitem [{\citenamefont {Bauer}\ \emph {et~al.}(2022)\citenamefont {Bauer}, \citenamefont {Neubert}, \citenamefont {Renner}, \citenamefont {Schnubel},\ and\ \citenamefont {Thamm}}]{Bauer:2021mvw}%
  \BibitemOpen
  \bibfield  {author} {\bibinfo {author} {\bibfnamefont {M.}~\bibnamefont {Bauer}}, \bibinfo {author} {\bibfnamefont {M.}~\bibnamefont {Neubert}}, \bibinfo {author} {\bibfnamefont {S.}~\bibnamefont {Renner}}, \bibinfo {author} {\bibfnamefont {M.}~\bibnamefont {Schnubel}}, \ and\ \bibinfo {author} {\bibfnamefont {A.}~\bibnamefont {Thamm}},\ }\href {\doibase 10.1007/JHEP09(2022)056} {\bibfield  {journal} {\bibinfo  {journal} {JHEP}\ }\textbf {\bibinfo {volume} {09}},\ \bibinfo {pages} {056} (\bibinfo {year} {2022})},\ \Eprint {http://arxiv.org/abs/2110.10698} {arXiv:2110.10698 [hep-ph]} \BibitemShut {NoStop}%
\bibitem [{\citenamefont {Ayala}\ \emph {et~al.}(2014)\citenamefont {Ayala}, \citenamefont {Dom{\'\i}nguez}, \citenamefont {Giannotti}, \citenamefont {Mirizzi},\ and\ \citenamefont {Straniero}}]{Ayala:2014pea}%
  \BibitemOpen
  \bibfield  {author} {\bibinfo {author} {\bibfnamefont {A.}~\bibnamefont {Ayala}}, \bibinfo {author} {\bibfnamefont {I.}~\bibnamefont {Dom{\'\i}nguez}}, \bibinfo {author} {\bibfnamefont {M.}~\bibnamefont {Giannotti}}, \bibinfo {author} {\bibfnamefont {A.}~\bibnamefont {Mirizzi}}, \ and\ \bibinfo {author} {\bibfnamefont {O.}~\bibnamefont {Straniero}},\ }\href {\doibase 10.1103/PhysRevLett.113.191302} {\bibfield  {journal} {\bibinfo  {journal} {Phys. Rev. Lett.}\ }\textbf {\bibinfo {volume} {113}},\ \bibinfo {pages} {191302} (\bibinfo {year} {2014})},\ \Eprint {http://arxiv.org/abs/1406.6053} {arXiv:1406.6053 [astro-ph.SR]} \BibitemShut {NoStop}%
\bibitem [{\citenamefont {Dolan}\ \emph {et~al.}(2022)\citenamefont {Dolan}, \citenamefont {Hiskens},\ and\ \citenamefont {Volkas}}]{Dolan:2022kul}%
  \BibitemOpen
  \bibfield  {author} {\bibinfo {author} {\bibfnamefont {M.~J.}\ \bibnamefont {Dolan}}, \bibinfo {author} {\bibfnamefont {F.~J.}\ \bibnamefont {Hiskens}}, \ and\ \bibinfo {author} {\bibfnamefont {R.~R.}\ \bibnamefont {Volkas}},\ }\href {\doibase 10.1088/1475-7516/2022/10/096} {\bibfield  {journal} {\bibinfo  {journal} {JCAP}\ }\textbf {\bibinfo {volume} {10}},\ \bibinfo {pages} {096} (\bibinfo {year} {2022})},\ \Eprint {http://arxiv.org/abs/2207.03102} {arXiv:2207.03102 [hep-ph]} \BibitemShut {NoStop}%
\bibitem [{\citenamefont {Caputo}\ and\ \citenamefont {Raffelt}(2024)}]{Caputo:2024oqc}%
  \BibitemOpen
  \bibfield  {author} {\bibinfo {author} {\bibfnamefont {A.}~\bibnamefont {Caputo}}\ and\ \bibinfo {author} {\bibfnamefont {G.}~\bibnamefont {Raffelt}},\ }\href {\doibase 10.22323/1.454.0041} {\bibfield  {journal} {\bibinfo  {journal} {PoS}\ }\textbf {\bibinfo {volume} {COSMICWISPers}},\ \bibinfo {pages} {041} (\bibinfo {year} {2024})},\ \Eprint {http://arxiv.org/abs/2401.13728} {arXiv:2401.13728 [hep-ph]} \BibitemShut {NoStop}%
\bibitem [{\citenamefont {Anastassopoulos}\ \emph {et~al.}(2017)\citenamefont {Anastassopoulos} \emph {et~al.}}]{CAST:2017uph}%
  \BibitemOpen
  \bibfield  {author} {\bibinfo {author} {\bibfnamefont {V.}~\bibnamefont {Anastassopoulos}} \emph {et~al.} (\bibinfo {collaboration} {CAST}),\ }\href {\doibase 10.1038/nphys4109} {\bibfield  {journal} {\bibinfo  {journal} {Nature Phys.}\ }\textbf {\bibinfo {volume} {13}},\ \bibinfo {pages} {584} (\bibinfo {year} {2017})},\ \Eprint {http://arxiv.org/abs/1705.02290} {arXiv:1705.02290 [hep-ex]} \BibitemShut {NoStop}%
\bibitem [{\citenamefont {Armengaud}\ \emph {et~al.}(2019)\citenamefont {Armengaud} \emph {et~al.}}]{IAXO:2019mpb}%
  \BibitemOpen
  \bibfield  {author} {\bibinfo {author} {\bibfnamefont {E.}~\bibnamefont {Armengaud}} \emph {et~al.} (\bibinfo {collaboration} {IAXO}),\ }\href {\doibase 10.1088/1475-7516/2019/06/047} {\bibfield  {journal} {\bibinfo  {journal} {JCAP}\ }\textbf {\bibinfo {volume} {06}},\ \bibinfo {pages} {047} (\bibinfo {year} {2019})},\ \Eprint {http://arxiv.org/abs/1904.09155} {arXiv:1904.09155 [hep-ph]} \BibitemShut {NoStop}%
\bibitem [{\citenamefont {Aaij}\ \emph {et~al.}(2015)\citenamefont {Aaij} \emph {et~al.}}]{LHCb:2015nkv}%
  \BibitemOpen
  \bibfield  {author} {\bibinfo {author} {\bibfnamefont {R.}~\bibnamefont {Aaij}} \emph {et~al.} (\bibinfo {collaboration} {LHCb}),\ }\href {\doibase 10.1103/PhysRevLett.115.161802} {\bibfield  {journal} {\bibinfo  {journal} {Phys. Rev. Lett.}\ }\textbf {\bibinfo {volume} {115}},\ \bibinfo {pages} {161802} (\bibinfo {year} {2015})},\ \Eprint {http://arxiv.org/abs/1508.04094} {arXiv:1508.04094 [hep-ex]} \BibitemShut {NoStop}%
\bibitem [{\citenamefont {Aaij}\ \emph {et~al.}(2017)\citenamefont {Aaij} \emph {et~al.}}]{LHCb:2016awg}%
  \BibitemOpen
  \bibfield  {author} {\bibinfo {author} {\bibfnamefont {R.}~\bibnamefont {Aaij}} \emph {et~al.} (\bibinfo {collaboration} {LHCb}),\ }\href {\doibase 10.1103/PhysRevD.95.071101} {\bibfield  {journal} {\bibinfo  {journal} {Phys. Rev. D}\ }\textbf {\bibinfo {volume} {95}},\ \bibinfo {pages} {071101} (\bibinfo {year} {2017})},\ \Eprint {http://arxiv.org/abs/1612.07818} {arXiv:1612.07818 [hep-ex]} \BibitemShut {NoStop}%
\bibitem [{\citenamefont {Anelli}\ \emph {et~al.}(2015)\citenamefont {Anelli} \emph {et~al.}}]{SHiP:2015vad}%
  \BibitemOpen
  \bibfield  {author} {\bibinfo {author} {\bibfnamefont {M.}~\bibnamefont {Anelli}} \emph {et~al.} (\bibinfo {collaboration} {SHiP}),\ }\href@noop {} {\  (\bibinfo {year} {2015})},\ \Eprint {http://arxiv.org/abs/1504.04956} {arXiv:1504.04956 [physics.ins-det]} \BibitemShut {NoStop}%
\bibitem [{\citenamefont {Ovchynnikov}\ and\ \citenamefont {Zaporozhchenko}(2025)}]{Ovchynnikov:2025gpx}%
  \BibitemOpen
  \bibfield  {author} {\bibinfo {author} {\bibfnamefont {M.}~\bibnamefont {Ovchynnikov}}\ and\ \bibinfo {author} {\bibfnamefont {A.}~\bibnamefont {Zaporozhchenko}},\ }\href {\doibase 10.1103/568p-d1ls} {\bibfield  {journal} {\bibinfo  {journal} {Phys. Rev. D}\ }\textbf {\bibinfo {volume} {112}},\ \bibinfo {pages} {015001} (\bibinfo {year} {2025})},\ \Eprint {http://arxiv.org/abs/2501.04525} {arXiv:2501.04525 [hep-ph]} \BibitemShut {NoStop}%
\bibitem [{\citenamefont {Curtin}\ \emph {et~al.}(2019)\citenamefont {Curtin} \emph {et~al.}}]{Curtin:2018mvb}%
  \BibitemOpen
  \bibfield  {author} {\bibinfo {author} {\bibfnamefont {D.}~\bibnamefont {Curtin}} \emph {et~al.},\ }\href {\doibase 10.1088/1361-6633/ab28d6} {\bibfield  {journal} {\bibinfo  {journal} {Rept. Prog. Phys.}\ }\textbf {\bibinfo {volume} {82}},\ \bibinfo {pages} {116201} (\bibinfo {year} {2019})},\ \Eprint {http://arxiv.org/abs/1806.07396} {arXiv:1806.07396 [hep-ph]} \BibitemShut {NoStop}%
\bibitem [{\citenamefont {O'Hare}(2021)}]{OHare:2021utq}%
  \BibitemOpen
  \bibfield  {author} {\bibinfo {author} {\bibfnamefont {C.~A.~J.}\ \bibnamefont {O'Hare}},\ }\href {\doibase 10.1103/PhysRevLett.127.251802} {\bibfield  {journal} {\bibinfo  {journal} {Phys. Rev. Lett.}\ }\textbf {\bibinfo {volume} {127}},\ \bibinfo {pages} {251802} (\bibinfo {year} {2021})},\ \Eprint {http://arxiv.org/abs/2109.03116} {arXiv:2109.03116 [hep-ph]} \BibitemShut {NoStop}%
\bibitem [{\citenamefont {Barr}(2012)}]{Barr:2011cz}%
  \BibitemOpen
  \bibfield  {author} {\bibinfo {author} {\bibfnamefont {S.~M.}\ \bibnamefont {Barr}},\ }\href {\doibase 10.1103/PhysRevD.85.013001} {\bibfield  {journal} {\bibinfo  {journal} {Phys. Rev. D}\ }\textbf {\bibinfo {volume} {85}},\ \bibinfo {pages} {013001} (\bibinfo {year} {2012})},\ \Eprint {http://arxiv.org/abs/1109.2562} {arXiv:1109.2562 [hep-ph]} \BibitemShut {NoStop}%
\bibitem [{\citenamefont {Barr}\ and\ \citenamefont {Chen}(2013)}]{Barr:2013tea}%
  \BibitemOpen
  \bibfield  {author} {\bibinfo {author} {\bibfnamefont {S.~M.}\ \bibnamefont {Barr}}\ and\ \bibinfo {author} {\bibfnamefont {H.-Y.}\ \bibnamefont {Chen}},\ }\href {\doibase 10.1007/JHEP10(2013)129} {\bibfield  {journal} {\bibinfo  {journal} {JHEP}\ }\textbf {\bibinfo {volume} {10}},\ \bibinfo {pages} {129} (\bibinfo {year} {2013})},\ \Eprint {http://arxiv.org/abs/1309.0020} {arXiv:1309.0020 [hep-ph]} \BibitemShut {NoStop}%
\bibitem [{\citenamefont {Barr}\ and\ \citenamefont {Scherrer}(2016)}]{Barr:2015lya}%
  \BibitemOpen
  \bibfield  {author} {\bibinfo {author} {\bibfnamefont {S.~M.}\ \bibnamefont {Barr}}\ and\ \bibinfo {author} {\bibfnamefont {R.~J.}\ \bibnamefont {Scherrer}},\ }\href {\doibase 10.1088/1475-7516/2016/05/065} {\bibfield  {journal} {\bibinfo  {journal} {JCAP}\ }\textbf {\bibinfo {volume} {05}},\ \bibinfo {pages} {065} (\bibinfo {year} {2016})},\ \Eprint {http://arxiv.org/abs/1508.07469} {arXiv:1508.07469 [hep-ph]} \BibitemShut {NoStop}%
\bibitem [{\citenamefont {Martin}(1993)}]{Martin:1992mj}%
  \BibitemOpen
  \bibfield  {author} {\bibinfo {author} {\bibfnamefont {S.~P.}\ \bibnamefont {Martin}},\ }\href {\doibase 10.1016/0550-3213(93)90114-5} {\bibfield  {journal} {\bibinfo  {journal} {Nucl. Phys. B}\ }\textbf {\bibinfo {volume} {398}},\ \bibinfo {pages} {359} (\bibinfo {year} {1993})},\ \Eprint {http://arxiv.org/abs/hep-ph/9211292} {arXiv:hep-ph/9211292} \BibitemShut {NoStop}%
\bibitem [{\citenamefont {Arkani-Hamed}\ \emph {et~al.}(2000)\citenamefont {Arkani-Hamed}, \citenamefont {Cheng}, \citenamefont {Dobrescu},\ and\ \citenamefont {Hall}}]{Arkani-Hamed:2000ulw}%
  \BibitemOpen
  \bibfield  {author} {\bibinfo {author} {\bibfnamefont {N.}~\bibnamefont {Arkani-Hamed}}, \bibinfo {author} {\bibfnamefont {H.-C.}\ \bibnamefont {Cheng}}, \bibinfo {author} {\bibfnamefont {B.~A.}\ \bibnamefont {Dobrescu}}, \ and\ \bibinfo {author} {\bibfnamefont {L.~J.}\ \bibnamefont {Hall}},\ }\href {\doibase 10.1103/PhysRevD.62.096006} {\bibfield  {journal} {\bibinfo  {journal} {Phys. Rev. D}\ }\textbf {\bibinfo {volume} {62}},\ \bibinfo {pages} {096006} (\bibinfo {year} {2000})},\ \Eprint {http://arxiv.org/abs/hep-ph/0006238} {arXiv:hep-ph/0006238} \BibitemShut {NoStop}%
\bibitem [{\citenamefont {Arkani-Hamed}\ \emph {et~al.}(2001)\citenamefont {Arkani-Hamed}, \citenamefont {Cohen},\ and\ \citenamefont {Georgi}}]{ArkaniHamed:2001nc}%
  \BibitemOpen
  \bibfield  {author} {\bibinfo {author} {\bibfnamefont {N.}~\bibnamefont {Arkani-Hamed}}, \bibinfo {author} {\bibfnamefont {A.~G.}\ \bibnamefont {Cohen}}, \ and\ \bibinfo {author} {\bibfnamefont {H.}~\bibnamefont {Georgi}},\ }\href {\doibase 10.1016/S0370-2693(01)00741-9} {\bibfield  {journal} {\bibinfo  {journal} {Phys. Lett.}\ }\textbf {\bibinfo {volume} {B513}},\ \bibinfo {pages} {232} (\bibinfo {year} {2001})},\ \Eprint {http://arxiv.org/abs/hep-ph/0105239} {arXiv:hep-ph/0105239 [hep-ph]} \BibitemShut {NoStop}%
\bibitem [{\citenamefont {Arkani-Hamed}\ \emph {et~al.}(2002{\natexlab{a}})\citenamefont {Arkani-Hamed}, \citenamefont {Cohen}, \citenamefont {Katz}, \citenamefont {Nelson}, \citenamefont {Gregoire},\ and\ \citenamefont {Wacker}}]{ArkaniHamed:2002qx}%
  \BibitemOpen
  \bibfield  {author} {\bibinfo {author} {\bibfnamefont {N.}~\bibnamefont {Arkani-Hamed}}, \bibinfo {author} {\bibfnamefont {A.~G.}\ \bibnamefont {Cohen}}, \bibinfo {author} {\bibfnamefont {E.}~\bibnamefont {Katz}}, \bibinfo {author} {\bibfnamefont {A.~E.}\ \bibnamefont {Nelson}}, \bibinfo {author} {\bibfnamefont {T.}~\bibnamefont {Gregoire}}, \ and\ \bibinfo {author} {\bibfnamefont {J.~G.}\ \bibnamefont {Wacker}},\ }\href {\doibase 10.1088/1126-6708/2002/08/021} {\bibfield  {journal} {\bibinfo  {journal} {JHEP}\ }\textbf {\bibinfo {volume} {08}},\ \bibinfo {pages} {021} (\bibinfo {year} {2002}{\natexlab{a}})},\ \Eprint {http://arxiv.org/abs/hep-ph/0206020} {arXiv:hep-ph/0206020 [hep-ph]} \BibitemShut {NoStop}%
\bibitem [{\citenamefont {Arkani-Hamed}\ \emph {et~al.}(2002{\natexlab{b}})\citenamefont {Arkani-Hamed}, \citenamefont {Cohen}, \citenamefont {Katz},\ and\ \citenamefont {Nelson}}]{ArkaniHamed:2002qy}%
  \BibitemOpen
  \bibfield  {author} {\bibinfo {author} {\bibfnamefont {N.}~\bibnamefont {Arkani-Hamed}}, \bibinfo {author} {\bibfnamefont {A.~G.}\ \bibnamefont {Cohen}}, \bibinfo {author} {\bibfnamefont {E.}~\bibnamefont {Katz}}, \ and\ \bibinfo {author} {\bibfnamefont {A.~E.}\ \bibnamefont {Nelson}},\ }\href {\doibase 10.1088/1126-6708/2002/07/034} {\bibfield  {journal} {\bibinfo  {journal} {JHEP}\ }\textbf {\bibinfo {volume} {07}},\ \bibinfo {pages} {034} (\bibinfo {year} {2002}{\natexlab{b}})},\ \Eprint {http://arxiv.org/abs/hep-ph/0206021} {arXiv:hep-ph/0206021 [hep-ph]} \BibitemShut {NoStop}%
\bibitem [{\citenamefont {Hook}(2018)}]{Hook:2018jle}%
  \BibitemOpen
  \bibfield  {author} {\bibinfo {author} {\bibfnamefont {A.}~\bibnamefont {Hook}},\ }\href {\doibase 10.1103/PhysRevLett.120.261802} {\bibfield  {journal} {\bibinfo  {journal} {Phys. Rev. Lett.}\ }\textbf {\bibinfo {volume} {120}},\ \bibinfo {pages} {261802} (\bibinfo {year} {2018})},\ \Eprint {http://arxiv.org/abs/1802.10093} {arXiv:1802.10093 [hep-ph]} \BibitemShut {NoStop}%
\bibitem [{\citenamefont {Cs\'aki}\ \emph {et~al.}(2018)\citenamefont {Cs\'aki}, \citenamefont {Geller},\ and\ \citenamefont {Telem}}]{Csaki:2017eio}%
  \BibitemOpen
  \bibfield  {author} {\bibinfo {author} {\bibfnamefont {C.}~\bibnamefont {Cs\'aki}}, \bibinfo {author} {\bibfnamefont {M.}~\bibnamefont {Geller}}, \ and\ \bibinfo {author} {\bibfnamefont {O.}~\bibnamefont {Telem}},\ }\href {\doibase 10.1007/JHEP05(2018)134} {\bibfield  {journal} {\bibinfo  {journal} {JHEP}\ }\textbf {\bibinfo {volume} {05}},\ \bibinfo {pages} {134} (\bibinfo {year} {2018})},\ \Eprint {http://arxiv.org/abs/1710.08921} {arXiv:1710.08921 [hep-ph]} \BibitemShut {NoStop}%
\bibitem [{\citenamefont {Cs\'aki}\ \emph {et~al.}(2020)\citenamefont {Cs\'aki}, \citenamefont {Guan}, \citenamefont {Ma},\ and\ \citenamefont {Shu}}]{Csaki:2019coc}%
  \BibitemOpen
  \bibfield  {author} {\bibinfo {author} {\bibfnamefont {C.}~\bibnamefont {Cs\'aki}}, \bibinfo {author} {\bibfnamefont {C.-S.}\ \bibnamefont {Guan}}, \bibinfo {author} {\bibfnamefont {T.}~\bibnamefont {Ma}}, \ and\ \bibinfo {author} {\bibfnamefont {J.}~\bibnamefont {Shu}},\ }\href {\doibase 10.1103/PhysRevLett.124.251801} {\bibfield  {journal} {\bibinfo  {journal} {Phys. Rev. Lett.}\ }\textbf {\bibinfo {volume} {124}},\ \bibinfo {pages} {251801} (\bibinfo {year} {2020})},\ \Eprint {http://arxiv.org/abs/1904.03191} {arXiv:1904.03191 [hep-ph]} \BibitemShut {NoStop}%
\bibitem [{\citenamefont {Cheng}\ and\ \citenamefont {Chung}(2020)}]{Cheng:2020dum}%
  \BibitemOpen
  \bibfield  {author} {\bibinfo {author} {\bibfnamefont {H.-C.}\ \bibnamefont {Cheng}}\ and\ \bibinfo {author} {\bibfnamefont {Y.}~\bibnamefont {Chung}},\ }\href {\doibase 10.1007/JHEP10(2020)175} {\bibfield  {journal} {\bibinfo  {journal} {JHEP}\ }\textbf {\bibinfo {volume} {10}},\ \bibinfo {pages} {175} (\bibinfo {year} {2020})},\ \Eprint {http://arxiv.org/abs/2007.11780} {arXiv:2007.11780 [hep-ph]} \BibitemShut {NoStop}%
\end{thebibliography}%

\end{document}